\documentclass[12pt]{article}
\input amssym.def
\input amssym
\newcommand{\be}{\begin{equation}}
\newcommand{\ee}{\end{equation}}
\begin{document}

\renewcommand{\theequation}{\arabic{section}.\arabic{equation}}

\title{Discrete model of spacetime in terms of inverse spectra
of the $T_0$ Alexandroff topological spaces}
\author{Vladimir N. Efremov\thanks{Mathematics Department, CUCEI,
University of Guadalajara, Guadalajara, Jalisco, Mexico.}
\thanks{Postal address: Apartado Postal 1-2011, C.P. 44100,
Guadalajara, Jalisco, M\'exico. E-mail:
efremov@udgserv.cencar.udg.mx}\\
and\\
Nikolai V. Mitskievich\thanks{Physics Department, CUCEI,
University of Guadalajara, Guadalajara, Jalisco, M\'exico.}
\thanks{Postal address: Apartado Postal 1-2011, C.P. 44100,
Guadalajara, Jalisco, M\'exico. E-mail:
nmitskie@udgserv.cencar.udg.mx}}
\date{~}

\maketitle

\noindent {\bf Running title:}  $T_0$-discrete spacetime
and inverse spectra

\newpage

\begin{abstract}
The theory of inverse spectra of $T_0$ Alexandroff topological
spaces is used to construct a model of $T_0$-discrete
four-dimensional spacetime. The universe evolution is interpreted
in terms of a sequence of topology changes in the set of
$T_0$-discrete spaces realized as nerves of the canonical
partitions of three-dimensional compact manifolds. The
cosmological time arrow arises being connected with the refinement
of the canonical partitions, and it is defined by the action of
homomorphisms in the proper inverse spectrum of three-dimensional
$T_0$-discrete spaces. A new causal order relation in this
spectrum is postulated having the basic properties of the causal
order in the pseudo-Riemannian spacetime however also bearing
certain quasi-quantum features. An attempt is made to describe
topological changes between compact manifolds in terms of
bifurcations of proper inverse spectra; this led us to the concept
of bispectrum. As a generalization of this concept, inverse
multispectra and superspectrum are introduced. The last one
enables us to introduce the discrete superspace, a discrete
counterpart of the Wheeler--DeWitt superspace.
\end{abstract}

~~~~~~~~~~~\\


\newpage
\section{Introduction}  \label{s1}

Absolutization of any concept (here: in physics), although it is
inevitable at certain stages of the development of the theory,
later always leads to a contradiction (Bohm, 1965). Exactly this
occurred with the concept of a smooth spacetime manifold. This
concept, being successful in the classical relativistic physics as
a model of the causally ordered set at large scales, under
extrapolation to the quantum theory leads to appearance of the
well known singularities and divergences. This is primarily
connected with the ideal character of pointlike events and objects
which can be recorded by classical observers (by their
definition). The standard quantum theory, from its very beginning,
has changed the approach to the definition of an observer and
observables, however leaving as untouchable the concept of smooth
spacetime manifold. But it seems to be natural and aesthetically
attractive to accompany the quantization of physical fields by
quantifying the spacetime arena, or even perform the latter in
anticipation, the arena on (or together with) which evolve these
fields. The idea to use dicrete (finitary) structures as
fundamental and really existing, but not approximational ones, in
the description of the quantum spacetime relations, dates back to
the works of Finkelstein (1969, 1988) and Isham (1989). Sorkin and
co-authors have proposed both finitary substitutes to model causal
relations between events (spacetime causal sets) in realistic
measurements (Bombelli {\em et al.}, 1987) and finitary
topological structures to model the quantized spacetime (Sorkin,
1991). Recently these ideas were extensively developed in (Rideout
and Sorkin, 2000; Raptis, 2000a; Raptis and Zapatrin, 2001;
Mallios and Raptis, 2001).

Nor one has to absolutize the concrete discrete spacetime
relations, not only since at the classical level ( some `large
scales') the smooth spacetime manifold does describe the
corresponding physical reality adequately, but, more importantly,
since the concrete discreteness differs drastically at different
levels. When passing to a deeper structural level, one has to be
ready to discover that objects previously treated as `elementary',
should be considered as compound ones, `built' of the next-level
`elementary' objects (remember, for example, the fate of hadrons
later interpreted via quarks). From our point of view, the sound
mathematical concept which describes both the discreteness and
continuity ideas, as well as their interconnection, is the inverse
spectrum of three-dimensional $T_0$-discrete spaces, also known as
the $T_0$ Alexandroff spaces (Alexandroff, 1937, 1947; Arenas,
1997, 1999). This inverse spectrum has as its limit the continuous
three-dimensional space (usually, image of a standard spacelike
section of the spacetime), but this continuous space is never
reached in the spectral evolution process. In this connection note
that our approach differs from that of Sorkin and his co-authors
as well as his followers; it is more similar to the approach of
Isham (1989), namely to the canonical Hamiltonian description
where the discretization is immediately applied to the
three-dimensional space, but not to the full four-dimensional
spacetime. However we do not take the 3-space as a section of the
latter, but consider its spectral evolution in the course of acts
of refinement which are inevitably also discrete, thus giving
birth to a new (discrete) parameter, the `global time'. Thus, in
our opinion, the spacetime is modelled by the proper inverse
spectrum of three-dimensional $T_0$ Alexandroff spaces, while the
global discrete evolution (the time arrow in the expanding
universe) is related to a sequential refinement of the canonical
partitions of a three-dimensional compact. Thus the global time
automatically acquires the $T_0$-discrete topology, since the
family of canonical partitions is a partially ordered infinite set
(Alexandroff, 1937). This spectral evolution parameter yields only
one additional dimension (timelike, see Subsection \ref{s2.5}
where the concept of light cone is introduced without metrization)
to the $n$ spatial dimensions (here, three) postulated from the
very beginning. See also the $(n+1)$-argumentation in the
framework of the conventional quantum theory given by van Dam and
Ng (2001).

Our model can be related to the (3+1)-splitting of spacetime into
a family of three-dimensional spacelike hypersurfaces
(equivalently, to introduction of a normal congruence of timelike
worldlines of local observers). This representation of the
four-dimensional spacetime continuum is used in the canonical
formulation of general relativity in terms of observables [see
(Misner {\em et al.}, 1973; Ashtekar, 1991) and references
therein] since it presumes introduction of a reference frame as a
continual system of observers situated at all points of the
three-dimensional spacelike hypersurface and moving along the
respective lines of the congruence. This method is also known as
the monad formalism (giving a covariant description of global
reference frames), see (Mitskievich, 1996), and its application to
the canonical formulation of general relativity, (Antonov {\em et
al.}, 1978). Thus the discretization of three-dimensional
hypersurfaces via a transition to the nerves of finite or locally
finite coverings (in particular, partitions), automatically leads
to a finite (for compacts or for compact regions of paracompacts)
set of observers, and to a denumerable set of events they can
detect in the evolving universe, {\em i.e.} in the course of
shifting along the inverse spectrum in the direction of
progressive refinement of the coverings.

In Section \ref{s2}, we give a review of the basic mathematical
concepts such as the $T_0$-discrete space ($T_0$ Alexandroff
space), nerves of coverings (partitions) and of inverse spectra of
topological spaces associated with the nerves. Moreover,
Alexandroff's procedure of discretization of compacts
(construction of the proper inverse spectrum of any compact) is
described, the procedure which is also applicable to paracompacts.
These items are fairly well known to theoretical physicists after
the paper (Sorkin, 1991) [see also (Raptis and Zapatrin, 2001;
Mallios and Raptis, 2001)], but we present them in a style closer
to the original papers of Alexandroff (1929, 1937, 1947) less
accessible to the English-speaking reader; furthermore, this style
we use in the physical interpretation of our results in
Subsections \ref{s2.4}, \ref{s2.5}. In this connection it is worth
mentioning that our starting attitude (discretization of
three-dimensional spacelike sections) does not give us the
possibility to reinterpret the obtained $T_0$-discrete spaces in
terms of causal sets as this was done in (Sorkin, 1991; Raptis,
2000a) where the four-dimensional spacetime manifolds were
discretized. Therefore in Subsection \ref{s2.5} we postulate a new
causal order relation in the proper inverse spectrum $S_{{\rm
pr}}$ defining two sets, those of the causal past and causal
future of any element of $S_{{\rm pr}}$. Further we prove two
Propositions justifying this postulate.

In Section \ref{s3} there is made an attempt to describe
topological changes between compact manifolds in terms of
bifurcations of proper inverse spectra. This led us in Subsection
\ref{s3.2} to the concept of bispectrum. In Subsection \ref{s3.3}
the concepts of inverse multispectra and superspectrum are
introduced. In our opinion, this last concept should be the
discrete counterpart of the superspace of the Wheeler--DeWitt
quantum geometrodynamics. The introduction of these concepts makes
it possible to set the problem of formulation of the quantum
theoretical approach to the topodynamics (an analogue of
geometrodynamics), and to propose the topological version of the
many-worlds interpretation as well as a qualitative discrete-space
analogue of Heisenberg's uncertainty relation.

\section{Inverse spectra of the $T_0$ Alexandroff spaces
and their physical interpretation}  \label{s2}
\setcounter{equation}{0}

\subsection{$T_0$ Alexandroff spaces, partially ordered
sets and simplicial complexes}  \label{s2.1}

By the Alexandroff space we mean a topological space $D$ every
point of which has a minimal neighborhood or, equivalently, the
space has a unique minimal base (Alexandroff, 1937) (the minimal
neighborhood of a point $p\in D$ is denoted by $O(p)$ being the
intersection of all open sets containing $p$). This is also
equivalent to the fact that intersection of any family of open
sets is open, and union of any number of closed sets is closed.
Therefore for each Alexandroff space $D$, there is a dual space
$D^*$ in which open sets are by a definition the closed sets of
$D$, and {\em vice versa}.

We consider here only the Alexandroff spaces with the $T_0$
separability axiom [of any given two points of a topological space
$D$, at least one is contained in an open set not containing the
other point (Hocking and Young, 1988)]. Note that an Alexandroff
space $D$ is $T_1$ iff $O(p)=p$ for any $p\in D$; in this case the
space $D$ is trivially discrete (discrete in the common sense).
But if we accept the $T_0$ axiom, a richer concept of discreteness
arises for which there exists a functorial equivalence between the
categories of $T_0$ Alexandroff spaces and partially ordered sets
(hereafter referred to as {\em posets}). We shall use as synonyms
`$T_0$ Alexandroff space' and `$T_0$-discrete space' [following
Alexandroff (1937): ``Diskrete R\"aume''], while the discrete
spaces in the common sense will be called `$T_1$-discrete spaces'
as well. Given a $T_0$ Alexandroff space $D$, we construct a poset
$P(D)$ with the order $p'\leqslant p$ iff $p\in O(p')$.
Conversely, given a poset $P$, we construct $T_0$ Alexandroff
space $D(P)$ with the topology generated by the minimal
neighborhoods \be  \label{2.1} O(p')=\{p\in P|p\geqslant p'\}. \ee
It is straightforward to see that $D(P(D))=D$ and $P(D(P))=P$ and
that under the functors, continuous mappings become order
preserving mapping and conversely (Arenas, 1997). Note that the
order can be also defined in the reversed way and we obtain the
$T_0$-discrete space $D^*$ dual to $D$: \be  \label{2.2}
O^*(p')=\{p\in P|p'\geqslant p\} \ee is the minimal neighborhood
of the point $p'$ in $D^*(P)$.

A $T_0$ Alexandroff space $D$ is locally finite if for any point
$p\in D$ the number of elements in $O(p)$ and in $\bar{p}$ is
finite. We denote as $\bar{p}$ the closure of the point $p\in D$.
The points with the property $\bar{p}=p$ are called $c$-vertices,
and those with the property $O(p)=p$, $o$-vertices.

Now let $V$ be a set of (abstract) elements called vertices. An
abstract simplicial complex $K$ is a collection of finite subsets
of $V$ with the property that each element of $V$ lies in some
element of $K$, and if $s$ is any element of $K$ (called simplex
of $K$), then any subset $s'$ of $s$ is again a simplex of $K$
($s'$ is said to be a face of $s$). In this case, if we suppose
that $s'\leqslant s$, the simplicial complex $K$ turns into the
poset $P(K)$, and therefore into the $T_0$-discrete space
$D(P(K))$, or it turns into the dual one, $D^*(P(K))$ [see
(\ref{2.1}) or (\ref{2.2})].

\subsection{Nerves of partitions and nerves' inverse spectra}
  \label{s2.2}

Nerves of coverings (in particular, canonical partitions) of
normal spaces represent an important example of (abstract)
simplicial complexes and $T_0$-discrete spaces.

Let $X$ be a normal topological space, {\em i.e.} a Hausdorff
space satisfying the $T_4$ separability axiom (Hocking and
Young,1988). A subset $A$ of the space $X$ is canonically closed
if $A$ is a closure of its interior $\dot{A}$, {\em i.e.}
$$
A=\bar{\dot{A ~}}.
$$
A canonical partition of the space $X$ is defined as a finite
covering consisting of canonically closed sets, \be  \label{2.3}
\alpha=\{A_1,\dots,A_s\}, \ee with disjoint interiors, {\em i.e}
$\dot{A}_i\cap \dot{A}_j=\emptyset$ for $\forall i,j=1,\dots s$;
$i\neq j$.

A canonical partition $\beta=\{B_1,\dots,B_r\}$ is called a
refinement of $\alpha$ if for any element $B_j\in\beta$ there is a
unique element $A_i\in\alpha$ such that $A_i$ contains $B_j$
($B_j\subseteq A_i$). It is worth being emphasized that, in the
case of partition, if such an element $A_i$ exists, it is
necessarily unique. It is also said that the partition $\beta$
follows $\alpha$ ($\beta\succ\alpha$).

For any pair $\alpha, ~ \beta$ of canonical partitions, there
exists a canonical partition $\gamma$ being a refinement of the
both $\alpha$ and $\beta$. The sets having this property are
called directed ones. Such a partition $\gamma$ may be obtained,
for example, as a product $\alpha\wedge\beta$ of the partitions
$\alpha$ and $\beta$ which consists of $\overline{\dot{A_i}\cap
\dot{B_j}}$ for all elements $A_i$ and $B_j$ for which
$\dot{A_i}\cap\dot{B_j}\neq\emptyset$. It is obvious that the
collection of all canonical partitions $\{\alpha\}$ of a normal
space $X$ is a partially ordered set, therefore $\{\alpha\}$ is a
{\em directed poset}.

Now, following Alexandroff (1937), we introduce a special case of
the $T_0$-discrete spaces which are realized as nerves of the
coverings of a normal space $X$.

Let $\alpha=\{A_1,\dots,A_s\}$ be a covering (in particular, a
canonical partition) of the normal space $X$. As a nerve of the
covering $\alpha$, we call the simplicial complex $N_\alpha$
consisting of simplices defined as sets
$\{A_{i_0},\dots,A_{i_q}\}$ of elements of the covering $\alpha$
for which \be  \label{2.4} A_{i_0}\cap\dots\cap
A_{i_q}\neq\emptyset. \ee It is said that the simplex
$s^q_\alpha=\{A_{i_0},\dots, A_{i_q}\}$ has the dimension $q$.

In accordance with the general procedure of determination of the
topology on a simplicial complex, one has to consider the simpices
$s^q_\alpha$ as points of the topological space and define the
minimal neighborhood of the simplex
$s^q_\alpha=\{A_{i_0},\dots,A_{i_q}\}$ as the set of simplices
$s^p_\alpha=\{A_{j_0},\dots,A_{j_p}\}$ such that \be   \label{2.5}
A_{i_0}\cap\dots\cap A_{i_q}\subseteq A_{j_0}\cap\dots\cap
A_{j_p}. \ee In other words, the minimal neighborhood
$O^*(s^q_\alpha)$ of the simplex $s^q_\alpha$ form all its faces
$s^p_\alpha$, {\em i.e.} \be  \label{2.6}
O^*(s^q_\alpha)=\left\{s^p_\alpha\in N_\alpha|s^q_\alpha \geqslant
s^p_\alpha\right\}. \ee Thus the $T_0$-discrete dual topology has
been defined on the nerve $N_\alpha$.

Exactly the nerves of canonical partitions with the $T_0$-discrete
dual topology are usually employed to the end of definition of
spectra of $T_0$-discrete spaces.

Let $\{\alpha\}$ be a set of coverings (canonical partitions) of a
normal space $X$, and $N_\alpha$, a nerve corresponding to a
covering $\alpha\in\{\alpha\}$; moreover, let $X_\alpha$ be a
$T_0$-discrete space defined on the basis of the nerve $N_\alpha$
via (\ref{2.5}) or (\ref{2.6}). The inverse spectrum of the nerves
$N_\alpha$ is defined as the set
$S=\{N_\alpha,\omega^{\alpha'}_\alpha \}$ where
$\omega^{\alpha'}_\alpha$ are simplicial mappings \be  \label{2.7}
\omega^{\alpha'}_\alpha: ~ N_{\alpha'}\rightarrow N_\alpha \ee
which are well defined only when $\alpha'$ is a refinement of
$\alpha$ ($\alpha'\succ\alpha$), while for $\alpha''\succ
\alpha'\succ\alpha$ the transitivity condition \be  \label{2.8}
\omega^{\alpha''}_\alpha=\omega^{\alpha'}_\alpha
\omega^{\alpha''}_{\alpha'} \ee should be fulfilled. (By the
definition, a simplicial mapping $\omega^{\alpha'}_\alpha$ maps
any simplex from $N_{\alpha'}$ in a simplex in $N_\alpha$.) The
inverse spectrum of $T_0$-discrete spaces $S=\{X_\alpha,
\omega^{\alpha'}_\alpha\}$ is introduced in the same manner, only
the mappings $\omega^{\alpha'}_\alpha$ now should be continuous.
Since we shall consider below nerves with the dual topology, the
pairs of objects, $N_\alpha$ and $X_\alpha$, will be identified
($N_\alpha \Longleftrightarrow X_\alpha$).

A point $\{s_\alpha\}$ of the direct product
$\displaystyle\prod_\alpha X_\alpha$ of the $T_0$-discrete spaces
corresponding to all coverings of the set $\{\alpha\}$, is called
a {\em coherent system} of elements $s_\alpha$ ({\em thread},
Alexandroff's term) of the inverse spectrum
$S=\{X_\alpha,\omega^{\alpha'}_\alpha\}$, if $s_\alpha=
\omega^{\alpha'}_\alpha s_{\alpha'}$ whenever
$\alpha'\succ\alpha$. The set of threads of a spectrum $S$
represents a subspace $\bar{S}$ of the topological space
$\displaystyle\prod_\alpha X_\alpha$. The subspace $\bar{S}$
endowed with the induced topology is called the {\em total inverse
limit} of the spectrum $S$, \be   \label{2.9}
\bar{S}=\lim\left\{X_\alpha,\omega^{\alpha'}_\alpha \right\}. \ee

For us however of greater importance will be the concept of the
{\em upper inverse limit}. First observe that a thread
$s=\{s_\alpha\}$ is larger than $\tilde{s}= \{\tilde{s}_\alpha\}$,
if for any $\alpha\in\{\alpha\}$, $s_\alpha\geq\tilde{s}_\alpha$
($\tilde{s}_\alpha$ is a face of $s_\alpha$). A thread $s$ is
called the maximum one, if no thread greater than $s$ exists. The
subspace $\hat{S}$ of the space $\bar{S}$ consisting of all
maximum threads, is called the upper inverse limit of the spectrum
$S$, \be  \label{2.10} \hat{S}={\rm
uplim}\left\{X_\alpha,\omega^{\alpha'}_\alpha \right\}. \ee

\subsection{Discretization of compacts}  \label{s2.3}

For all compacts there is a standard procedure how to construct
the upper inverse limit of the spectrum of $T_0$-discrete spaces
or of the nerves of all canonical partitions (finite by the
definition, see Subsection \ref{s2.2}) (Alexandroff, 1947). (For
paracompacts the situation is fairly similar, but finite
partitions should be changed to locally finite ones.)

Let $\{\alpha\}$ be a set of all canonical partitions of a compact
$X$. [First let us remark that the set $\{\alpha\}$ is cofinal to
the set of all open coverings of the compact $X$. This means that
for any open covering $\omega$ of $X$ there is a canonical
partition $\alpha_\omega\in\{\alpha\}$ being the refinement of
$\omega$ ($\alpha_\omega\succ \omega$).] Let us construct for a
canonical partition $\alpha=\{A_1,\dots,A_s\}$ the respective
nerve $N_\alpha$ and introduce on it the dual topology. This
yields a $T_0$-discrete space which we denote by $X_\alpha$. For
any partition $\alpha'=\{A'_1,\dots, A'_r\}$ being a refinement of
$\alpha$, the mapping
$\omega^{\alpha'}_\alpha:X_{\alpha'}\rightarrow X_\alpha$ is
determined as follows: to any $A'_j\in\alpha'$ there corresponds
only one $A_i\in \alpha$ which contains $A'_j$. This element $A_i$
of the partition $\alpha$ is by the definition the image of $A'_j$
under the mapping $\omega^{\alpha'}_\alpha$, {\em i.e.} \be
\label{2.11} \omega^{\alpha'}_\alpha A'_j=A_i. \ee Now for any
point $s^p_{\alpha'}=\{A'_{j_0}, \dots,A'_{j_p}\}\in X_{\alpha'}$
we have \be  \label{2.12} \omega^{\alpha'}_\alpha
s^p_{\alpha'}=\left\{ \omega^{\alpha'}_\alpha
A'_{j_0},\dots,\omega^{\alpha'}_\alpha
A'_{j_p}\right\}=\{A_{i_0},\dots,A_{i_q}\}=s^q_\alpha\in X_\alpha.
\ee We see that $q\leqslant p$ since among the sets
$\omega^{\alpha'}_\alpha A'_{j_0},\dots,\omega^{\alpha'}_\alpha
A'_{j_p}$ some sets may be the same. This construction completes
the deduction of the spectrum of $T_0$-discrete spaces; it is
called the {\em proper inverse spectrum} of the compact $X$
(Alexandroff, 1947), \be  \label{2.13} S_{{\rm
pr}}=\left\{X_\alpha, \omega^{\alpha'} _\alpha\right\}. \ee

Alexandroff (1947) has shown that any compact $X$ is homeomorphic
to the upper inverse limit of its proper spectrum $S_{{\rm pr}}$.
To prove this theorem he used the following realization of the
upper inverse limit of $S_{{\rm pr}}$.

For any point $x\in X$ there is a unique point \be  \label{2.14}
s^q_\alpha=\{A_{i_0},\dots,A_{i_q}\} \ee of the $T_0$-discrete
space $X_\alpha$ (the simplex of $N_\alpha$) such that $x\in
A_{i_0}\cap\dots\cap A_{i_q}$, but in the canonical partition
$\alpha$ there are no more sets which contain the point $x$. The
point $s^q_\alpha(x) \in X_\alpha$ is called the carrier of the
point $x$ in the discrete space $X_\alpha$.

The set $\left\{s^q_\alpha(x)\right\}$ of carriers of any point
$x$ in all spaces $X_\alpha$ satisfies the conditions
$\omega^{\alpha'}_\alpha s^q_{\alpha'}(x)=s^q_\alpha(x)$ whenever
$\alpha'\succ\alpha$, {\em i.e.} $\left\{ s^q_\alpha(x)\right\}$
is a thread of the proper spectrum $S_{{\rm pr}}$. It is easy to
show that any thread of the carriers
$\left\{s^q_\alpha(x)\right\}$ of any point $x\in X$ forms a
maximum thread. Therefore to any point $x\in X$ corresponds a
unique maximum thread $\left\{ s^q_\alpha(x)\right\}$ pertaining
to the upper inverse limit $\hat{S}_{{\rm pr}}$. The inverse
assertion is also true, namely that any maximum thread
$\{s_\alpha\}\in \hat{S}_{{\rm pr}}$ is a thread of carriers of a
certain point $x\in X$. Hence there exists a bijective mapping
$f:X\rightarrow\hat{S}_{{\rm pr}}$ between the compact $X$ and the
upper inverse limit $\hat{S}_{{\rm pr}}$ of its proper spectrum
$S_{{\rm pr}}$ since $s^q_\alpha(x) =s^q_\alpha(x')$ for all
canonical partitions $\alpha$ of the compact $X$ yields $x=x'$. If
in the space $\hat{S}_{{\rm pr}}$ has been introduced the topology
induced by the inverse total limit $\bar{S}_{{\rm pr}}$ (see
Subsection \ref{s2.2}), thus the bijective mapping $f$ becomes a
homeomorphism between $X$ and $\hat{S}_{ {\rm pr}}$.

\subsection{Physical interpretation of inverse spectra of
$T_0$ Ale\-xandroff spaces}  \label{s2.4}

Our model of spacetime is based on the assumption that the
fundamental (and existing as a reality) is considered the
$T_0$-discrete three-dimensional space, whose topology is evolving
in the induced $T_0$-discrete time, while a continuous  spatial
section of the spacetime is treated as a limiting
three-dimensional manifold which never is realized in the course
of this discrete evolution. From the results of Alexandroff
described in the preceding Subsection, it follows that for any
three-dimensional compact $X$ there exists at least one inverse
spectrum $S_{{\rm pr}}=\left\{X_\alpha,\omega^{\alpha'
}_\alpha\right\}$ whose upper limit is homeomorphic to the compact
$X$. We may treat this spectrum as a primary object describing the
discrete spacetime manifold (in the terminology of Riemann), while
the compact $X$ is merely a result of the limiting process. The
set of $T_0$ Alexandroff's spaces $X_\alpha$ in the inverse
spectrum $S_{{\rm pr}}$ then is interpreted as a family of
$T_0$-discrete analogues of three-dimensional sections of the
four-dimensional continuum $M$.

From the canonical approach to general relativity it is known that
to the end of description of the gravitational field in terms of
observables, it is necessary to split the spacetime manifold $M$
into a complete family of three-dimensional spacelike
hypersurfaces, introducing at the same time the congruence of
timelike worldlines of local observers orthogonal to this family.
The completeness of the family of spacelike hypersurfaces means
that through any event (worldpoint) $p\in M$ passes one and only
one hypersurface. Hence this family represents a linearly ordered
(one-parametric) set of three-dimensional spacelike sections. From
the viewpoint of the monad method (Mitskievich, 1996) (see more
references therein) this procedure of splitting spacetime manifold
$M$ is nothing but a choice of a (classical) reference frame, {\em
i.e.} of a multitude of local test observers whose worldlines are
identified with lines of the non-rotating congruence, while the
spacelike sections orthogonal to the congruence, are the
three-dimensional simultaneity hypersurfaces.

Returning to the construction of a discrete model of the spacetime
manifold $M$ we suppose that the role of three-dimensional
hypersurfaces of simultaneity is played namely by the $T_0$
Alexandroff spaces $X_\alpha$. Any two points $s^p_\alpha$ and
$s^q_\alpha$ of $T_0$-discrete space $X_\alpha$ ({\em i.e.}
simplices $s^p_\alpha$ and $s^q_\alpha$ of the nerve $N_\alpha$ of
the canonical partition $\alpha$) are interpreted as two
simultaneous events occurring on the $T_0$-discrete hypersurface
$X_\alpha$ at the instant of the $T_0$-discrete time labelled by
the partition $\alpha$.

\underline{Observation 2.1.} The proposed interpretation suggests
that the set of canonical partitions $\{\alpha\}$ should be
considered as the many-arrow time ["many-fingered" time in (Misner
{\em et al}, 1793) \S 21.8]. This time-labelling set (by virtue of
its partial orderedness) possesses a $T_0$-discrete topology (see
\ref{s2.1}). Now we accept the hypothesis: If
$\alpha'\succ\alpha$, the $T_0$-discrete space $X_{\alpha'}$ is in
the future with respect to $X_\alpha$. Then the homomorphism
$\omega^{\alpha'}_\alpha:X_{\alpha'}\rightarrow X_\alpha$ (defined
in the spectrum $S_{{\rm pr}}$ iff $\alpha'\succ \alpha$) plays
the role of a shift in the $T_0$-discrete time $\{\alpha\}$ from
the future to the past. It is worth emphasizing that the
homomorphism $\omega^{\alpha'}_\alpha$ determines a topological
transition between two non-homeomorphic $T_0$ Alexandroff spaces
$X_\alpha$ and $X_{\alpha'}$. Thus in this model a $T_0$-discrete
time shift always is a topology change.

In order to model the splitting of a spacetime manifold into a
complete family of spacelike hypersurfaces, {\em i.e.} to
construct the discrete analogue of the classical reference frame,
it is necessary to single out of a partially ordered set (poset)
$\{\alpha\}$, a {\em complete linearly ordered subset (closset)}
of canonical partitions $\{\alpha_i|i\in I\}$. Remember that the
set $\{\alpha_i|i\in I\}$ is called linearly ordered if for any
two different elements $\alpha_i$ and $\alpha_j$ is true either
$\alpha_i\succ\alpha_j$ or $\alpha_j\succ\alpha_i$. The {\em
completeness} of the subset $\{\alpha_i|i\in I\}$ we define as
follows: any element $\alpha^*\in\{\alpha\}$
such that\\
(a) $\alpha_i\succ\alpha^*\succ\alpha_j$ for some
pair of elements $\alpha_i,\alpha_j\in\{\alpha_i|i\in I\}$,\\
(b) for any $\alpha_i\in\{\alpha_i|i\in I\}$
is true either $\alpha_i\succ\alpha^*$
or $\alpha^*\succ\alpha_i$\\
pertains to $\{\alpha_i|i\in I\}$.

Now we shall prove that any closset $\{\alpha_i|i\in I\}$ of a
poset $\{\alpha\}$, which is cofinal to this poset, is
denumerable.

To this end let us consider an arbitrary closset $\{ \alpha_i|i\in
I\}$ cofinal to the poset $\{\alpha\}$. Due to its cofinality the
closset $\{\alpha_i|i\in I\}$ is infinite. If it is denumerable,
this is the end of the proof. Suppose that $\{\alpha_i|i\in I\}$
is non-denumerable and see that this leads to a contradiction.
Select out of the closset $\{\alpha_i|i\in I\}$ a denumerable
subset $\left\{\alpha_{i_k}|k\in{\Bbb Z}^+ \right\}$ (${\Bbb Z}^+$
is the set of nonnegative integer numbers) being cofinal to the
closset $\{\alpha_i|i\in I\}$. Such a subset always exists for any
compact (Alexandroff, 1947). This subset is linearly ordered due
to the linear orderedness of $\{\alpha_i|i\in I\}$. Consider any
pair of consecutive partitions $\alpha_{i_k}$ and
$\alpha_{i_{k+1}}$. Since $\alpha_{i_k}\prec\alpha_{ i_{k+1}}$,
and the both partitions are finite (by the definition of canonical
partitions, Subsection \ref{s2.1}), there exists only a {\em
finite} set of partitions
$\alpha_{i_k}(1),\dots,\alpha_{i_k}(C_k)$ out of the closset
$\{\alpha_i|i\in I\}$, such that
$$
\alpha_{i_k}\prec\alpha_{i_k}(1)\prec\dots\prec
\alpha_{i_k}(C_k)\prec\alpha_{i_{k+1}}.
$$
It is clear that the obtained set of partitions \be   \label{2.15}
\left\{\alpha_{i_k},\alpha_{i_k}(1),\dots,
\alpha_{i_k}(C_k)|k\in{\Bbb Z}^+\right\} \ee is equivalent to the
initial set $\{\alpha_i|i\in I\}$. But the set (\ref{2.15}) is
denumerable due to the fact that a set consisting of a denumerable
set of finite sets, is denumerable. Thus the closset
$\{\alpha_i|i\in I\}$ is denumerable, {\em i.e.} $I\cong{\Bbb
Z}^+$ ($I$ is equivalent to ${\Bbb Z}^+$ in the sense that the
both have one and the same cardinality).

Now we consider three particular cases of clossets of poset
$\{\alpha\}$.

(1) If a closset $\{\alpha_i|i\in I\}$ is cofinal to the poset
$\{\alpha\}$ and it includes the trivial partition
$\alpha_0=\{X\}$ consisting of one element (the compact $X$
proper), then the inverse subspectrum \be  \label{2.16} S_{{\rm
pr}} (\alpha_i ,i\in I)=\left\{X_{\alpha_i},
\omega^{\alpha_j}_{\alpha_i} |i,j\in I\right\} \ee of the inverse
spectrum $S_{{\rm pr}}$, describes a linearly ordered family of
$T_0$-discrete sections from the $T_0$-discrete spacetime
corresponding to $S_{{\rm pr}}$. This means that the inverse
spectrum $S_{{\rm pr}} (\alpha_i ,i\in I)$ should be considered as
a model of the $T_0$-discrete spacetime in a fixed reference
frame. To reiterate, the just introduced concept of discrete
reference frame includes a complete family of linearly ordered
(with a discrete time parameter $\{\alpha_i|i\in I\}$)
three-dimensional $T_0$ Alexandroff spaces and a system of
homomorphisms between any two $T_0$-discrete spaces in this
family,
$$
\omega^{\alpha_j}_{\alpha_i}:X_{\alpha_j}\rightarrow
X_{\alpha_i} ~ \mbox{ whenever } ~ \alpha_j\succ\alpha_i.
$$

\underline{Observation 2.2.} Subspectrum $S_{{\rm pr}} (\alpha_i
,i\in I)$, like the spectrum $S_{{\rm pr}}$, represents a proper
inverse spectrum whose upper limit is homeomorphic to the upper
limit of $S_{{\rm pr}}$
$$
{\rm uplim}\left\{X_{\alpha_i},\omega^{\alpha_j}_{\alpha_i}
\right\}\cong{\rm uplim}\left\{X_\alpha,
\omega^{\alpha'}_\alpha\right\}\cong X.
$$
Thus the inverse spectrum $S_{{\rm pr}}(\alpha_i ,i\in I)$ is
modelling the same four-dimensional spacetime continuum $M$, as
does the spectrum $S_{{\rm pr}}$, but now in a fixed reference
frame.

\underline{Observation 2.3.} It was observed by Sorkin (Sorkin,
1995) that macroscopic space volume is supposed to be measured by
the number of points on the $T_0$-discrete section $X_{\alpha_i}$
(the proposition adapted to our spacetime model). Then the
discrete evolution of the universe described by a complete
linearly ordered spectrum $S_{{\rm pr}}(\alpha_i,i\in I)$,
represents creation of the universe beginning with only one-point
space $X_{\alpha_0}$ ($\alpha_0=\{X\}$ is the trivial partition of
the three-space $X$), with the subsequent expansion (in the sense
of the growth of the number of points) in the course of
transitions between $T_0$-discrete hypersurfaces from
$X_{\alpha_i}$ to $X_{\alpha_j}$ where $\alpha_j\succ\alpha_i$.
The homomorphism
$\omega^{\alpha_j}_{\alpha_i}:X_{\alpha_j}\rightarrow
X_{\alpha_i}$ describes a topology change with a decrease of the
volume (in the above sense), being an operator acting in the
direction opposite to the flow of the global (cosmological) time
in the observed expanding universe.

(2) If a closset $\{\alpha_i|i\in I\}$ is cofinal to the poset
$\{\alpha\}$, but contains a minimal non-trivial partition
$\alpha_{{\rm min}}= \{A_1,\dots,A_s\}$, $s>1$, {\em i.e.} such
that for any $i\in I$, $\alpha_i\succcurlyeq\alpha_{ {\rm min}}$,
we shall say that the related inverse subspectrum
$$
S_{{\rm pr}}(\alpha_{{\rm min}})=\left\{X_{\alpha_i},
\omega^{\alpha_j}_{\alpha_i}|\alpha_i\succcurlyeq\alpha_{
{\rm min}},i,j\in I\right\}
$$
is modelling $T_0$-discrete spacetime (in a fixed reference
frame), but only beginning with the $T_0$-discrete time instant
labelled by $\alpha_{{\rm min}}$. In this case, like in (1), the
set of indices $I$ is equivalent to ${\Bbb Z}^+$.

(3) Alternatively, if the closset $\{\alpha_i|i\in I\}$ is not
cofinal to the poset $\{\alpha\}$, but contains both minimal and
maximal partitions $\alpha_{{\rm min}}$ and $\alpha_{{\rm max}}$,
such that $\alpha_{{\rm min}}
\preccurlyeq\alpha_i\preccurlyeq\alpha_{{\rm max}}$ for any $i\in
I$, then the inverse subspestrum
$$
S_{{\rm pr}}(\alpha_{{\rm min}},\alpha_{{\rm max}})=
\left\{X_{\alpha_i},\omega^{\alpha_j}_{\alpha_i}|
\alpha_{{\rm min}}\preccurlyeq\alpha_i\preccurlyeq
\alpha_{{\rm max}},i,j\in I\right\}
$$
of the spectrum $S_{{\rm pr}}$, describes a $T_0$-discrete
spacetime sandwich. This sandwich contains a finite number of
$T_0$-discrete hypersurfaces $X_{\alpha_i}$, since $\alpha_{{\rm
min}}$ and $\alpha_{{\rm max}}$ are finite partitions. This
$T_0$-discrete spacetime sandwich is an analogue of a finite
region of the spacetime continuum between two non-intersecting
spacelike hypersurfaces. Such sandwiches are used in the canonical
approach to general relativity when the Cauchy problem is
considered [see (Baierline {\em et al.}, 1962; Misner {\em et
al.}, 1973)].

Returning to the case (1), observe that from the poset
$\{\alpha\}$ of all canonical partitions of the space $X$, it is
possible to single out an infinite set of clossets everyone of
which corresponds to a certain discrete reference frame. For
example, let us consider closset $\{\alpha'_i|i\in I\}$ of poset
$\{\alpha\}$ cofinal to $\alpha$ and including the trivial
partition $\alpha'_0= \{X\}$. Let us also suppose that for all
indices $i\in I$, with the exception of $i=0$, the canonical
partitions $\alpha_i$ and $\alpha'_i$ are not ordered with respect
to $\succ$. In this case we shall say that the inverse spectra \be
\label{2.17}
\begin{array}{rcl}
S_{{\rm pr}} (\alpha_i ,i\in I)&=&\left\{X_{\alpha_i},
\omega^{\alpha_j}_{\alpha_i}|i,j\in I\right\},\\
S_{{\rm pr}} (\alpha'_i ,i\in I)&=&\left\{X_{\alpha'_i},
\omega^{\alpha'_j}_{\alpha'_i}|i,j\in I\right\}
\end{array}
\ee describe one and the same $T_0$-discrete spacetime, but in
different discrete reference frames due to the unorderedness of
the $T_0$-discrete sections $X_{\alpha_i}$ and $X_{\alpha'_i}$ for
all $i\in I$.

One can describe a transition between these reference frames,
taking a partition $\alpha''_i$ for any pair of unordered
partitions $\alpha_i$ and $\alpha'_i$, such that
$\alpha''_i\succ\alpha_i$, $\alpha''_i\succ\alpha'_i$. (Such a
partition $\alpha''_i$ does exist, {\em e.g.},
$\alpha''_i=\alpha_i\wedge\alpha'_i$, due to the directedness of
the set $\{\alpha\}$; however, $\alpha''_i$ may not pertain to any
of the clossets $\{\alpha_i|i\in I\}$ and $\{\alpha'_i|i\in I\}$.)
Then we have two homomorphisms
$$
\omega^{\alpha''_i}_{\alpha_i}:X_{\alpha''_i}\rightarrow
X_{\alpha_i} ~ \mbox{ and } ~ \omega^{\alpha''_i}_{\alpha'_i}
:X_{\alpha''_i}\rightarrow X_{\alpha'_i},
$$
and the transition from the first discrete reference frame,
$S_{{\rm pr}} (\alpha_i ,i\in I)$, to the second one, $S_{{\rm
pr}} (\alpha'_i ,i\in I)$, is described as a mapping \be
\label{2.18}
X_{\alpha'_i}=\omega^{\alpha''_i}_{\alpha'_i}\left(\omega^{
\alpha''_i}_{\alpha_i}\right)^{-1}X_{\alpha_i}. \ee Here
$\left(\omega^{\alpha''_i}_{\alpha_i}\right)^{-1}$ is the
many-valued mapping inverse to the homomorphism
$\omega^{\alpha''_i}_{\alpha_i}$. The many-valuedness of this
mapping is due to the property $\alpha''_i\succ\alpha_i$ which is
related to the very idea of description of the $T_0$-discrete
spacetime in terms of inverse spectra of the three-dimensional
$T_0$ Alexandroff spaces. This probably reflects the fact that the
concept of discrete reference frame introduced in our model,
acquires, due to spacetime discretization, certain quasi-quantum
properties. With an exceptional sharpness these features are
revealed in the construction of an analogue of the monad
description of the global reference frame (in classical general
relativity, as a congruence of worldlines of test observers). Now
instead of the congruence of worldlines of classical observers we
take the complete system of maximal threads being the upper limit
of the inverse spectrum.

If we associate the worldline of the observer with the maximal
thread $\{s_\alpha\}=\{s_\alpha\in X_\alpha|\alpha\in\{
\alpha\}\}$ of the proper inverse spectrum $S_{{\rm pr}}$, then
the events $s_\alpha$ on the discrete worldline of the `observer'
$\{s_\alpha\}$ will be partially ordered due to the partial
orderedness of the set $\{\alpha\}$ of all canonical partitions of
the compact $X$. In other words, the `observer' in this
interpretation exists in an infinite multitude of reference frames
at once, and the proper time of such an `observer' is actually
many-arrow time. However an observer in the classical relativistic
mechanics is fixing only one local reference frame (a single-arrow
time). Therefore a more adequate counterpart of observer's
worldline should be a {\em subthread} \be   \label{2.19}
\{s_{\alpha_i}\}=\{s_{\alpha_i}\in X_{\alpha_i}|\alpha_i\in\{
\alpha_i|i\in I\}\} \ee corresponding to the subspectrum $S_{{\rm
pr}}(\alpha_i,i\in I)$. (This exactly corresponds to the reference
frame concept introduced via a family of $T_0$-discrete
hypersurfaces $X_{\alpha_i}$.) In this case there is a linear
orderedness of the events $s_{\alpha_i}$ on the observer's
discrete worldline $\{s_{\alpha_i}\}$, meaning that any two
partitions $\alpha_i,\alpha_j\in\{\alpha_i|i\in I\}$ are ordered
(for example, $\alpha_j\succ\alpha_i$), thus $s_{\alpha_i}=
\omega^{\alpha_j}_{\alpha_i}s_{\alpha_j}$. However in the general
case through one point $s_{\alpha_i}\in X_{\alpha_i}$ goes not
one, but a finite or denumerable set of maximal threads of the
spectrum $S_{{\rm pr}}(\alpha_i,i\in I)$. Thus the upper limit
$\hat{S}_{{\rm pr}}(\alpha_i,i\in I)=
\mbox{uplim}\left\{X_{\alpha_i},\omega^{\alpha_j}_{\alpha_i}
\right\}$ defined as a complete system of maximal threads,
describes a set of ``multifurcating'' observers. [Observe that the
cardinality of a set of maximal threads (observers) is equal to
the cardinality of a three-dimensional compact $X$, the same which
is known for points on a spacelike hypersurface in general
relativity.] It is worth being emphasized that the furcations of
discrete worldlines of observers (threads) only occur in the
future direction, {\em i.e.} with transitions to more refined
partitions. This is the alternative expression of the fact that in
this model evolution of the expanding universe is described by a
sequence of topology changes (homomorphisms) within the class of
$T_0$ Alexandroff spaces \be  \label{2.20}
\omega^{\alpha_{i+1}}_{\alpha_i}:X_{\alpha_{i+1}}\rightarrow
X_{\alpha_i}, \ee as well as that the ``evolution operator''
$\left(\omega^{\alpha_{i+1}}_{\alpha_i}\right)^{-1}$ (inverse to
the homomorphism $\omega^{\alpha_{i+1}}_{\alpha_i}$) is
many-valued. These topology changes are in certain sense
quasi-quantum processes with respect to the smooth classical
evolution of three-geometries, {\em e.g.}, in the framework of the
canonical approach to general relativity.

\subsection{Establishment of the causal order in the
proper inverse spectrum} \label{s2.5}

In the Observation 2.1 a hypothesis was accepted which established
the partial ordering of three-dimensional $T_0$-discrete spatial
sections (if $\alpha'\succ\alpha$, the $T_0$-discrete space
$X_{\alpha'}$ is in future with respect to $X_\alpha$). This
brought us to the concept of a discrete reference frame (see
Observation 2.2). However the fact that two events $s_\alpha$ and
$s_{\alpha'}$ pertain to spaces $X_\alpha$ and $X_{\alpha'}$, does
not yet mean that the event $s_{\alpha'}$ is in the causal future
of the event $s_\alpha$. Just this situation takes place for the
3+1-splitting of the continuous spacetime in a family of
three-dimensional spacelike sections in the standard relativity
theory, and it would be natural to reproduce it in the discrete
case. This leads to the necessity to define in the proper inverse
spectrum $S_{{\rm pr}}$ (as a model of the $T_0$-discrete
spacetime) such a relation of partial ordering of events, which
would permit a causal interpretation, {\em i.e.} to define a
causal order in $S_{{\rm pr}}$.

To this end we first remark that from the definition of a complete
linearly ordered subset (closset) $\{\alpha_i|i\in I\}$ (see
Subsection 2.4) it follows that if a partition $\alpha_i$ contains
$N$ canonically closed sets, then $\alpha_{i+1}$ consists of $N+1$
sets, since otherwise a partition $\alpha* \in \{\alpha\}$ should
exist such that $\alpha_i\prec\alpha*\prec\alpha_{i+1}$, which
would not pertain to the closset $\{\alpha_i|i\in I\}$. This fact
contradicts to the supposition of completeness of $\{\alpha_i|i\in
I\}$. Thus we can define a {\em discrete time quantum} as a
transition from the $T_0$-discrete space $X_{\alpha_i}$ to
$X_{\alpha_{i+1}}$ for any $i\in I$.

Just for the three closest (in the discrete time sense) spaces
$X_{\alpha_{i-1}}, X_{\alpha_i}$ and $X_{\alpha_{i+1}}$, we
introduce the causal order between events, then we extend it
inductively to the inverse spectrum $S_{{\rm pr}}(\alpha_i, i\in
I)$ (\ref{2.16}), and finally to the whole proper inverse spectrum
$S_{{\rm pr}}$ (\ref{2.13}).

The causal past of an event $s_{\alpha_i}$ in the nearest past
space $X_{\alpha_{i-1}}$ we define as the set \be  \label{2.21}
\mbox{CP}(s_{\alpha_i})\cap X_{\alpha_{i-1}}\equiv \Lambda
(s_{\alpha_i})\cap X_{\alpha_{i-1}}
:=\overline{O(\omega^{\alpha_i}_{\alpha_{i-1}}
s_{\alpha_i})}\equiv \Lambda^i_{i-1} s_{\alpha_i}. \ee The just
defined operator $\Lambda^i_{i-1}$ can act both on a separate
point and on a subset of $X_{\alpha_i}$.

The verbal justification of this definition is that it is natural
to assume that with an event $s_{\alpha_i}$ are causally related
only such events $s_{\alpha_{i-1}}$ which pertain to the closure
of the minimal neighborhood
$O(\omega^{\alpha_i}_{\alpha_{i-1}}s_{\alpha_i})$ of the event
$\omega^{\alpha_i}_{\alpha_{i-1}}s_{\alpha_i}$ being the spectral
projection of the event $s_{\alpha_i}$ onto the $T_0$-discrete
space $X_{\alpha_{i-1}}$ which is the nearest past of
$X_{\alpha_i}$. This corresponds to some extent to the idea of
Finkelstein (1988) that the physical causal connection between
events in a discretized spacetime should be such that connects the
nearest-neighborhood events [see also (Raptis, 2000a)]. The
closure is taken in order to include the intersection of the space
$X_{\alpha_{i-1}}$ with the analogue of the past light cone of the
event $s_{\alpha_i}$.

For an event $s_{\alpha_i}$, its causal past on the space
$X_{\alpha_{i-k}}$ ($k$ is any positive integer) is obtained by
the $k$-fold action of operators of type $\Lambda^i_{i-1}$ with
successively decreasing indices \be  \label{2.22}
\mbox{CP}(s_{\alpha_i})\cap X_{\alpha_{i-k}}\equiv \Lambda
(s_{\alpha_i})\cap X_{\alpha_{i-k}} := \Lambda^{i-k+1}_{i-k}\cdots
\Lambda^i_{i-1} s_{\alpha_i}. \ee

Similarly to (\ref{2.21}), the causal future in the nearest future
space $X_{\alpha_{i+1}}$ of an event $s_{\alpha_i}$ is \be
\label{2.23} \mbox{CF}(s_{\alpha_i})\cap X_{\alpha_{i+1}}\equiv
V(s_{\alpha_i})\cap X_{\alpha_{i+1}}
:=(\omega^{\alpha_{i+1}}_{\alpha_i})^{-1}
\overline{Os_{\alpha_i}}\equiv V^i_{i+1}s_{\alpha_i}. \ee For an
event $s_{\alpha_i}$ its causal future on the space
$X_{\alpha_{i+k}}$ ($k$ is any positive integer) is \be
\label{2.24} \mbox{CF}(s_{\alpha_i})\cap X_{\alpha_{i+k}}\equiv
V(s_{\alpha_i})\cap X_{\alpha_{i+k}} := V^{i+k-1}_{i+k}\cdots
V^i_{i+1} s_{\alpha_i}. \ee This definition together with (2.22)
ensures the transitivity of the introduced causal order in the
proper inverse spectrum $S_{{\rm pr}}(\alpha_i, i\in I)$.

Passing to the introduction of the causal order in the inverse
spectrum $S_{{\rm pr}}$, we take two arbitrary canonical
partitions $\alpha$ and $\alpha'\in\{\alpha\}$ with only one
condition $\alpha'\succ\alpha$. Now consider the intersection of
the causal past $\mbox{CP}(s_{\alpha'})$ of any event
$s_\alpha'\in X_\alpha'$ with the space $X_\alpha$ and the
intersection of the causal future $\mbox{CF}(s_{\alpha})$ of any
event $s_\alpha\in X_\alpha$ with the space $X_\alpha'$. To this
end we define a closset of canonical partitions \be  \label{2.25}
\mbox{Cl}=\{\alpha_i|i\in\overline{0,f}, f\in\Bbb{Z}^+\} \ee such
that $\alpha_0 = \alpha, \alpha_f = \alpha'$ and $\alpha_i\succ
\alpha_{i-1}$ for $i\in\overline{1,f}$. Then the definitions
(2.22) and (2.24) yield \be  \label{2.26}
\mbox{CP}(s_{\alpha'})\cap X_{\alpha}\equiv \Lambda
(s_{\alpha'})\cap X_{\alpha} := \Lambda^{1}_{0}\dots
\Lambda^f_{f-1} s_{\alpha'}. \ee and \be  \label{2.27}
\mbox{CF}(s_{\alpha})\cap X_{\alpha'}\equiv V (s_{\alpha})\cap
X_{\alpha'} := V^{f-1}_{f}\dots V^0_1 s_{\alpha}. \ee Now we say
that the event $s_\alpha$ is in the causal past with respect to
the event $s_{\alpha'}$ if $s_\alpha\in\Lambda (s_{\alpha'})\cap
X_{\alpha}$ and we write $s_\alpha\prec s_{\alpha'}$. Also we
consider the event $s_{\alpha'}$ to be in the causal future with
respect to the event $s_\alpha$ if $s_{\alpha'}\in
V(s_{\alpha})\cap X_{\alpha'}$ and we write $s_{\alpha'}\succ
s_\alpha$. Immediately from these definitions follows the
transitivity of these relations, {\em i.e.}\\
\hspace*{2.cm} if $s_\alpha\prec s_{\alpha'}$ and
$s_\alpha'\prec s_{\alpha''}$,
then $s_\alpha\prec s_{\alpha''}$;\\
\hspace*{2.cm} if $s_{\alpha''}\succ s_{\alpha'}$ and
$s_{\alpha'}\succ s_\alpha$, then $s_{\alpha''}\succ s_\alpha$.

In the continuous pseudo-Riemannian spacetime $M^4$ these two
relations are equivalent, {\em i.e.} $s_{\alpha'}\succ s_\alpha$
implies $s_\alpha\prec s_{\alpha'}$ and {\em vice versa}. We prove
this equivalence for the discrete case in the framework of our
definitions. By virtue of the transitivity it is sufficient to
prove it for only one step, {\em i.e.} for the transition from
$\alpha_i$ to $\alpha_{i+1}$.

\underline{Proposition 2.1.} Let \be  \label{2.28}
\tilde{s}_{\alpha_{i+1}}\forall\in V(s_{\alpha_i})\cap
X_{\alpha_{i+1}}
=\left(\omega^{\alpha_{i+1}}_{\alpha_i}\right)^{-1}
\overline{Os_{\alpha_i}}, \ee then \be  \label{2.29}
s_{\alpha_i}\in\Lambda(\tilde{s}_{\alpha_{i+1}}) \cap
X_\alpha=\overline{O(\omega^{\alpha_{i+1}
}_{\alpha_i}\tilde{s}_{\alpha_{i+1}})}, \ee for any $X_{\alpha_i}$
and $X_{\alpha_{i+1}}$ such that $\alpha_{i+1}\succ{\alpha_i}$.

{\em Proof.} From (\ref{2.28}) it follows that
$\omega^{\alpha_{i+1}}_{\alpha_i}\tilde{s}_{\alpha_{i+1}}
\in\omega^{\alpha_{i+1}}_{\alpha_i}\left(
\omega^{\alpha_{i+1}}_{\alpha_i}\right)^{-1}
\overline{Os_{\alpha_i}}\subseteq\overline{Os_{\alpha_i}}$.
Setting $\tilde{s}_{\alpha_i}=
\omega^{\alpha_{i+1}}_{\alpha_i}\tilde{s}_{\alpha_{i+1}}$, we get
$\tilde{s}_{\alpha_i}\in\overline{Os_{\alpha_i}}$. We now have to
proof that $s_{\alpha_i}\in\overline{O \tilde{s}_{\alpha_i}}$. For
a $T_0$ Alexandroff space, it follows from
$\tilde{s}_{\alpha_i}\in\overline{O s_{\alpha_i}}$ that $\exists ~
s^*_{\alpha_i}\in O s_{\alpha_i}$ such that
$\tilde{s}_{\alpha_i}\in \overline{s^*_{\alpha_i}}$; then
$s^*_{\alpha_i}\in O\tilde{s}_{\alpha_i}$ [see (Alexandroff,
1937)], so that $\overline{s^*_{\alpha_i}}\subseteq\overline{O
\tilde{s}_{\alpha_i}}$. But from $s^*_{\alpha_i}\in Os_{\alpha_i}$
it also follows that $s_{\alpha_i}\in\overline{s^*_{\alpha_i}}$
(Alexandroff, 1937). Finally,
$s_{\alpha_i}\in\overline{s^*_{\alpha_i}}
\subseteq\overline{O\tilde{s}_{\alpha_i}}$. Q.E.D.

\underline{Proposition 2.2.} Let \be  \label{2.30}
s_{\alpha_i}\forall\in\Lambda(\tilde{s}_{\alpha_{i+1}}) \cap
X_{\alpha_i}= \overline{O(\omega^{\alpha_{i+1}}_{\alpha_i}
\tilde{s}_{\alpha_{i+1}})}, \ee then \be  \label{2.31}
\tilde{s}_{\alpha_{i+1}}\in V(s_{\alpha_i})\cap X_{\alpha_{i+1}}
=\left(\omega^{\alpha_{i+1}}_{\alpha_i}\right)^{-1}
\overline{Os_{\alpha_i}} \ee for any $X_{\alpha_i}$ and
$X_{\alpha_{i+1}}$ such that $\alpha_{i+1}\succ{\alpha_i}$.

{\em Proof.} With
$\tilde{s}_{\alpha_i}=\omega^{\alpha_{i+1}}_{\alpha_i}
\tilde{s}_{\alpha_{i+1}}$, (\ref{2.30}) takes the form
$s_{\alpha_i}\in\overline{O\tilde{s}_{\alpha_i}}$ yielding
$\tilde{s}_{\alpha_i}\in\overline{Os_{\alpha_i}}$ (see the proof
of the Proposition 2.1), {\em i.e.}
$\omega^{\alpha_{i+1}}_{\alpha_i}
\tilde{s}_{\alpha_{i+1}}\in\overline{Os_{\alpha_i}}$. Applying the
operator $\left(\omega^{\alpha_{i+1}}_{\alpha_i}\right)^{-1}$ to
the last relation, we get
$\left(\omega^{\alpha_{i+1}}_{\alpha_i}\right)^{-1}
\omega^{\alpha_{i+1}}_{\alpha_i}\tilde{s}_{\alpha_{i+1}}
\subseteq\left(\omega^{\alpha_{i+1}}_{\alpha_i}\right)^{-1}
\overline{Os_{\alpha_i}}$. Since $\tilde{s}_{\alpha_{i+1}}\in
\left(\omega^{\alpha_{i+1}}_{\alpha_i}\right)^{-1}
\omega^{\alpha_{i+1}}_{\alpha_i}\tilde{s}_{\alpha_{i+1}}$, we come
to (\ref{2.31}). This completes the proof.

The Propositions 2.1 and 2.2 show that the postulated causal
structure in the $T_0$-discrete spacetime modelled by the proper
inverse spectrum $S_{{\rm pr}}$, reveals the crucial properties of
the causal order in pseudo-Riemannian manifolds. In other words,
we have introduced the partial order corresponding to the
classical relativistic definition of the past and future.

\underline{Observation 2.4.} It is worth noting that the complete
linearly ordered subset (closset) of canonical partitions
(\ref{2.25}) between $\alpha$ and $\alpha'$ is not uniquely
defined. This is the result of the possibility to perform the
subdivision of sets of the partition $\alpha$ in different
successions, but finally yielding one and the same partition
$\alpha'$. [In the continuous spacetime, this corresponds to
consideration in a spacetime sandwich of different intermediate
hypersurfaces (different reference frames), while the initial and
final hypersurfaces remain fixed.] Due to the finiteness of the
partitions $\alpha$ and $\alpha'$, only a finite number of such
clossets $\mbox{Cl}_n$, $n\in\overline{1,N}$ exists. It is easy to
show on simple examples that different causal pasts $\Lambda
(s_{\alpha'})\cap X_{\alpha}$ and futures $V (s_{\alpha})\cap
X_{\alpha'}$ correspond to different choices of these
$\mbox{Cl}_n$, {\em cf.} (\ref{2.26}) and (\ref{2.27}). In the
framework of the concepts introduced above (see Subsection
\ref{s2.4}), this means that the causal past and future do depend
on the choice of discrete reference frame. This ambiguity of the
discrete light cone could be a manifestation of the {\em
quasi-quantum nature} of the proposed spacetime model. In the
contemporary universe such an ambiguity in the causal ordering for
different reference frames should be expected (according to our
model) at the near-discreteness (say, Planck) scales. Thus at the
resolution power of the present measurements the light cone should
be invariant under changes of reference frames. However, at the
near-discreteness scale resolution power, the concrete choice of
reference frame [which is an idealization of measurements, in this
case performed on essentially quantum objects and by quantum
observers, thus the act of measurement (see the first treatment of
this problem by Bohr and Rosenfeld (1933) in the framework of
quantum electrodynamics) merely becomes a specific kind of
interaction between two quantum objects] should influence these
objects so much that the classical concept of the test reference
body has to be abandoned. The discrete reference frame is itself a
set of intrinsically quantum objects, thus observer(s) and
observable(s) have now equal footing (the quasi-quantum nature
mentioned above).

\section{Multispectrum of an ensemble of compact
manifolds} \label{s3}
\setcounter{equation}{0}

\subsection{Prime summands and nice canonical partitions
of three-dimensional compact manifolds}  \label{s3.1}

In Section \ref{s2}, while constructing the proper inverse
spectrum of the compact $X$, we used canonical partitions of $X$,
{\em i.e.} such finite closed partitions $\alpha$ whose elements
have disjoint interiors. However we did not suppose the elements
of canonical partitions to be homeomorphic to three-dimensional
closed discs $\overline{D^3}$. This means that elements of a
canonical partition could have a rather complicated topological
structure. But in this case the nerve of a given canonical
partition does not fully reflect the topological complexity of the
compact $X$. Moreover, by a refinement of canonical partitions
their nerves yield a more exhaustive description of the compact's
topological structure. In this Section we are trying to construct
such a representation of any compact manifold $M$ and such a
sequence of its canonical partitions that the effect of
manifestation (in the corresponding sequence of nerves) of
topological structures at finer scales would become more explicit.

To this end we shall need some additional concepts [see {\em e.g.}
(Fomenko and Matveev, 1991)]

From each of two three-dimensional manifolds $M_1$ and $M_2$, an
open three-dimensional disk is removed, and the remainders are
glued together by means of a certain homeomorphism of the boundary
spheres. The obtained manifold $M_1\# M_2$ is called the {\em
connected sum} of $M_1$ and $M_2$. Note that the sphere $S^3$ is a
neutral element with respect to the connected sum operation in the
sense that $M\# S^3=S^3\# M=M$ for any three-manifold $M$. A
three-manifold is called {\em prime} if it cannot be represented
as a connected sum of two manifolds neither of which is a
three-sphere. Note that here, as in the number theory, neither
$S^3$ nor the number 1 are respectively considered as primes.

Our basic construction of the $h$-levels compact manifold
(hc-manifold) $M_h$ will follow from the Prime Decomposition
Theorem of Kneser (1929) which states that any three-dimensional
manifold $M$ can be represented as a finite connected sum of prime
manifolds. Milnor (1962) has proven that the prime summands of $M$
are uniquely determined by $M$ up to a homeomorphism.

Moreover, we shall use the so-called nice coverings and nice
partitions of compact manifolds. Remember that an open covering
$\omega=\{O_i|i\in I_\omega\}$ of $M$ is called {\em nice}, if all
nonempty intersections $O_{i_0}\cap\dots\cap O_{i_p}$ are
homeomorphic to the open disk $D^3$ (Bott and Tu, 1982). It is
known that any manifold possesses a nice covering. If the manifold
is compact, its nice covering can be chosen to be finite.

In order to define a nice canonical partition the Lebesgue lemma
would be helpful; it will be important also in determination of
the scales hierarchy on the hc-manifold $M_h$. Denoting an open
neighborhood of the set $A_i\subset M$ as $O(A_i)$, we may define
the {\em Lebesgue blowing up} of the partition $\alpha=\{A_i|i\in
I_\alpha\}$ as an open covering $\omega_\alpha=\{O(A_i)|i\in
I_\alpha\}$ having the nerve $N_{\omega_\alpha}$ isomorphic to the
nerve $N_\alpha$ of the partition $\alpha$. The Lebesgue lemma
states that the Lebesgue blowing up of any finite partition of a
compact always exists. We shall give this lemma in Alexandroff's
formulation (Alexandroff, 1998):\\
\underline{Lemma}. For any canonical partition $\alpha =\{A_i|i\in
I_\alpha\}$ of the compact $X$, one can find such a number
$l_\alpha>0$ called the {\em Lebesgue number} of the partition
$\alpha$, that the nerve $N_{\omega_\alpha}$ of the open covering
$\omega_\alpha=\{O(A_i,l_\alpha)|i\in I_\alpha\}$, is isomorphic
to the nerve $N_\alpha$ of the initial partition $\alpha$. Here
$O(A_i,l_\alpha)$ is the $l_\alpha$-neighborhood (the Lebesgue
neighborhood) of the set $A_i$, {\em i.e.}
$$
O(A_i,l_\alpha)=\{x\in X|\rho(x,y)<l_\alpha,y\forall\in A_i\},
$$
$\rho(x,y)$ being the distance between the points $x$ and $y$
of the compact $X$. (Compacts are normal spaces with denumerable
base, thus they are always metrizable.)

Let us call a {\em nice} canonical partition of a compact manifold
$M$, such one that has at least one Lebesgue blowing up which is a
nice open covering of $M$. Note that any set $A_i$ of a nice
canonical partition is homeomorphic to a closed disk
$\overline{D^3}$. Nice canonical partitions (in particular,
triangulations) describe topology of compact manifolds better than
all other coverings and partitions. More precisely, the \v{C}ech
homologies (and cohomologies) over nerves of the nice canonical
partitions of a compact manifold $M$ with coefficients in ${\Bbb
Z}$ are mutually isomorphic, being as well isomorphic to the
\v{C}ech homologies (cohomologies) of the very manifold $M$ (Bott
and Tu, 1982):
$$
\check{H}_*(N_\alpha,{\Bbb Z})\cong\check{H}_*(M,{\Bbb Z}), ~
~ \check{H}^*(N_\alpha,{\Bbb Z})\cong\check{H}^*(M,{\Bbb Z}).
$$
In other words, $T_0$ Alexandroff space $M_\alpha$ corresponding
to any nice canonical partition $\alpha$, fully describes
homological properties of the compact manifold $M$.

\subsection{Many-level compact manifolds
and description of topology changes}  \label{s3.2}

In this Subsection we construct a family of $h$-levels compact
manifolds $M_h$ for any $h\in{\Bbb Z}^+$, and describe the
sequence of topology changes between manifolds of different
levels, in terms of homomorphisms between the respective $T_0$
Alexandroff spaces.

We begin constructing with some closed compact connected manifold
$M_0$ which is supposed to be either prime or a sphere $S^3$.
First consider its trivial partition $\alpha_{-1}=\{M_0\}$
consisting only of the manifold $M_0$ itself. The nerve
$N_{\alpha_{-1}}$ of this partition consists of only one point.
This nerve by no means reflects the topology of $M_0$. Therefore
one has to introduce a nice canonical partition \be  \label{3.1}
\alpha_0=\{A_{i_0}|i_0\in I_0\} \ee where $I_0$ is a finite set of
indices. We shall call $M_0$ the 0-level compact manifold
(0c-manifold); let us for the convenience represent it as \be
\label{3.2} M_0=\bigcup_{i_0\in I_0}A_{i_0} \ee and characterize
it by the nerve $N_{\alpha_0}$ with a $T_0$-discrete topology
({\em i.e.} by a $T_0$ Alexandroff space $M_{0,\alpha_0}$).

In order to construct a 1c-manifold $M_1$ we introduce a
collection of manifolds \be   \label{3.3}
\sigma_1=\left\{\Sigma^1_k|k\in K_1\right\} \ee which includes the
sphere $S^3$ ($\Sigma^1_0=S^3$) as well as the prime compact
closed manifolds. ($K_1$ is a finite set of indices.) The question
which prime closed manifolds are included in this collection,
still remains open. Next we choose for each $i_0$ some
$\Sigma^1_{k(i_0)}\in\sigma_1$ and form the connected sum
$A_{i_0}\#\Sigma^1_{k(i_0)}$. Then the 1c-manifold $M_1$ is
determined by analogy with $M_0$ as \be  \label{3.4}
M_1=\bigcup_{i_0\in I_0}\left(A_{i_0}\#\Sigma^1_{k(i_0)} \right),
~ ~ k(i_0)\in K_1. \ee

\underline{Observation 3.1.} Actually, this definition gives a
family of 1c-mani\-folds which depend on the choice of the
integer-valued function $k=k(i_0)$, so that the exact notation for
the manifold (\ref{3.4}) should be $M_{1,k(i_0)}$. This family
contains $|K_1|^{|I_0|}$ $T_0$-discrete manifolds where $|K_1|$
and $|I_0|$ are numbers of elements in the sets $K_1$ and $I_0$
respectively. Thus when the function $k=k(i_0)$ is a fixed one, we
shall for the sake of conciseness adhere to the notation
(\ref{3.4}). If $k(i_0)=0$ for all $i_0\in I_0$, the manifold
remains unchanged, $M_1=M_0$.

Thus on $M_1$ is induced a canonical (though, in general, not
nice) partition \be  \label{3.5}
\alpha'_0=\left\{A'_{i_0}=A_{i_0}\#\Sigma^1_{k(i_0)}|i_0\in
I_0,k(i_0)\in K_1\right\}. \ee By virtue of the definition of the
partition $\alpha'_0$ its nerve is isomorphic to the nerve of
partition $\alpha_0$, $N_{\alpha'_0}\cong N_{\alpha_0}$, though
these nerves correspond to partitions of different manifolds,
$M_1$ and $M_0$ respectively. (This construction can be considered
as an example of the fact that $T_0$ Alexandroff spaces related to
different manifolds, $M_0$ and $M_1$, may be mutually
homeomorphic, $M_{0,\alpha_0}\cong M_{1,\alpha'_0}$.) It is worth
emphasizing that $\alpha_0$ is a nice partition, while $\alpha'_0$
is not.

To the end of a better representation of the topological structure
of the manifold $M_1$, let us introduce a nice partition
$\alpha_1$ of the manifold $M_1$ \be  \label{3.6}
\alpha_1=\{A_{i_0i_1}|i_0\in I_0,i_1\in I_1(i_0)\}, \ee more
refined than $\alpha'_0$. Here $I_1(\alpha_0)$ is a finite set of
indices which depends of $i_0$, such that \be  \label{3.7}
A'_{i_0}=\bigcup_{i_1\in I_1(i_0)}A_{i_0i_1}. \ee

\underline{Observation 3.2.} We impose an additional restriction
on diameters of the partitions $\alpha_0$, $\alpha'_0$, and
$\alpha_1$ which enables to compare diameters of partitions of
different manifolds (here, $M_0$ and $M_1$), and to speak whether
$\alpha_1$ really is a substantial refinement of $\alpha'_0$.
First note that the construction process of hc-manifolds can be
realized in the seven-dimensional Euclidean space ${\Bbb R}^7$,
therefore the diameters of partitions of all manifolds $M_h$
($h\in{\Bbb Z}^+$) can be compared in this universal embedding
space (Kodama, 1958). Suppose that pasting prime manifolds into
sets of the partition $\alpha_0$ does not alter their diameters,
{\em i.e.} $d(A_{i_0})=d(A_{i_0}\#\Sigma^1_{k(i_0)})$. This means
that diameters of the partitions $\alpha_0$ and $\alpha'_0$ in the
embedding space ${\Bbb R}^7$ are equal, \be  \label{3.8}
d(\alpha_0)=d(\alpha'_0). \ee On the other hand, we suppose that
\be  \label{3.9}
d(\alpha_1)\leqslant\min\left(\frac{1}{2}d(\alpha'_0),
\frac{1}{2}l(\alpha'_0)\right) \ee where $l(\alpha'_0)$ is the
Lebesgue number of the partition $\alpha'_0$. In this case we
shall call the partition $\alpha_1$, a {\em Lebesgue refinement}
of the partition $\alpha'_0$. This requirement [usual in the
description of  refinement of partitions to the end of creation of
the proper inverse spectrum (Alexandroff, 1929; Boltyansky, 1951)]
establishes a hierarchy between different levels of the manifold
$M_1$, which are described (by our definition) by the partitions
$\alpha'_0$ and $\alpha_1$ (or, similarly, between manifolds of
different levels, $M_0$ and $M_1$). Really, let us blow up the
sets of the partition $\alpha'_0$ by the diameter of the partition
$\alpha_1$, {\em i.e.} determine the open covering \be
\label{3.10}
\omega'_0=\left\{O\left(A'_{i_0},d(\alpha_1)\right)|i_0\in
I_0\right\}. \ee Then due to the condition (\ref{3.9}) we have the
homeomorphism of $T_0$-discrete spaces $M_{1,\alpha'_0} \cong
M_{1,\omega'_0}$. In other words, the $T_0$-discrete space
$M_{1,\omega'_0}$ is invariant with respect to fluctuations of the
diameter of the covering $\omega'_0$ in the limits determined by
the diameter $d(\alpha_1)$ of the next-level partition $\alpha_1$
of the manifold $M_1$. This means that any level of the refinement
in fact is represented by a whole class of coverings to which
correspond mutually isomorphic nerves. Thus there is a hierarchy
each level of which is exactly the above mentioned class; the
first step of the establishment of this hierarchy was described by
the condition (\ref{3.9}).

Returning to the expression (\ref{3.7}), we can determine the
homomorphism of the nerves $\omega^{\alpha_1}_{
\alpha'_0}:N_{\alpha_1}\rightarrow N_{\alpha'_0}$ and the
respective $T_0$ Alexandroff spaces \be  \label{3.11}
\omega^{\alpha_1}_{\alpha'_0}:M_{1,\alpha_1}\rightarrow
M_{1,\alpha'_0}. \ee Remembering the homeomorphism
$M_{0,\alpha_0}\cong M_{1,\alpha'_0}$, one can say that the
homomorphism $\omega^{\alpha_1}_{\alpha'_0}$ describes the
topology change between the discrete versions of the manifolds
$M_1$ and $M_0$, \be  \label{3.12}
\omega^{\alpha_1}_{\alpha'_0}:M_{1,\alpha_1}\rightarrow
M_{1,\alpha'_0}\cong M_{0,\alpha_0}. \ee

$T_0$-discrete space $M^{{\rm cr}}_{01}:= M_{0,\alpha_0}\cong
M_{1,\alpha'_0}$ may be called the critical level in the proper
inverse spectra of manifolds $M_0$ and $M_1$. One can say that at
the level of the $T_0$-discrete space $M^{{\rm cr}}_{01}$ occurs a
{\em bifurcation} of the inverse spectra of manifolds $M_0$ and
$M_1$ in the following sense: Introduce two sequences of
partitions $\{\alpha_{0,i}\}$ and $\{\alpha_{1,i}\}$ of the
manifolds $M_0$ and $M_1$ respectively. Let the index $i$ take
integer values from some $p_0$, to $+\infty$ ($i\in
\overline{p_0,\infty}$). For $i=p_0$ let $\alpha_{0,p_0}=
\{M_0\}$, $\alpha_{1,p_0}=\{M_1\}$, {\em i.e.} they are trivial
partitions to which correspond trivial homeomorphic $T_0$-discrete
spaces $M_{0,\alpha_{0,p_0}}\cong M_{1, \alpha_{1,p_0}}$. For
$i=p_1>p_0$ it is natural to take $\alpha_{0,p_1}=\alpha_0$ and
$\alpha_{1,p_1}=\alpha'_0$, thus the $T_0$-discrete spaces
corresponding to these partitions are also homeomorphic,
$M_{0,\alpha_{0,p_1}} \cong M_{1,\alpha_{1,p_1}}$, by virtue of
(\ref{3.12}). One may say that topology of the manifold $M_0$ is
equivalent to topology of $M_1$ at the scales $d>d(\alpha_0)
=d(\alpha'_0)$, see (\ref{3.8}). More precisely, there exist two
sequences of partitions ({\em enlargements} of the
partitions\footnote{A partition $\beta=\{B_j\}$ is called
enlargement of a partition $\alpha=\{A_i\}$ if each $B_j$ is a
union of some sets $A_i\in\alpha$ and each $A_i$ enters only one
union as a summand.} $\alpha_{0,p_1}$ and $\alpha_{1,p_1}$
respectively) \be  \label{3.13} \left.\begin{array}{l}
\alpha_{0,p_0}\prec\alpha_{0,p_0+1}\prec\dots\prec
\alpha_{0,p_1}=\alpha_0,\\
\alpha_{1,p_0}\prec\alpha_{1,p_0+1}\prec\dots\prec
\alpha_{1,p_1}=\alpha'_0 \end{array}\right\} \ee of the manifolds
$M_0$ and $M_1$, for which the $T_0$-discrete spaces are pairwise
homeomorphic, \be  \label{3.14} M_{0,\alpha_{0,p_0}}\cong
M_{1,\alpha_{1,p_0}}, ~ M_{0,\alpha_{0,p_0+1}}\cong
M_{1,\alpha_{1,p_0+1}}, \dots,M_{0,\alpha_{0,p_1}}\cong
M_{1,\alpha_{1,p_1}}. \ee It is however obvious that for $i>p_1$
the homeomorphism does not hold due to the definition of the
manifold $M_1$ through $M_0$ (\ref{3.4}). Thus the inverse spectra
\be  \label{3.15}
\left\{M_{0,\alpha_{0,i}},\omega^{\alpha_{0,i+1}}_{
\alpha_{0,i}}|i\in\overline{p_0,\infty}\right\}, ~ ~
\left\{M_{1,\alpha_{1,i}},\omega^{\alpha_{1,i+1}}_{
\alpha_{1,i}}|i\in\overline{p_0,\infty}\right\} \ee contain
homeomorphic $T_0$-discrete spaces for $i\in \overline{p_0,p_1}$
and no homeomorphic spaces for $i\in \overline{p_1+1,\infty}$.
Just due to the mentioned homeomorphism of the initial
$T_0$-discrete spaces in the spectra (\ref{3.15}), as well as due
to the fact that the manifold $M_1$ appears only at the level of
the partition $\alpha_{1,p_1}$, one may initiate the second
spectrum in (\ref{3.15}) with $i=p_1$. This yields the pair of
spectra, \be  \label{3.16}
\left\{M_{0,\alpha_{0,i}},\omega^{\alpha_{0,i+1}}_{
\alpha_{0,i}}|i\in\overline{p_0,\infty}\right\}, ~ ~
\left\{M_{1,\alpha_{1,i}},\omega^{\alpha_{1,i+1}}_{
\alpha_{1,i}}|i\in\overline{p_1,\infty}\right\}, \ee which we
shall call a {\em bispectrum}. We also shall say that the
$T_0$-discrete space $M^{{\rm cr}}_{01}:=
M_{0,\alpha_{0,p_1}}\cong M_{1,\alpha_{1,p_1}}$ represents the
critical level (the bifurcation level) of the bispectrum
(\ref{3.16}).

When the function $k(i_0)$ is not fixed ({\em i.e.} when all such
functions are taken into account), see the Observation 3.1, there
is a multifurcation at the critical level (see Subsection
\ref{s3.3}).

To construct the 2nd-level compact manifold (2c-manifold) $M_2$,
we take, generally speaking, another [in comparison with
(\ref{3.3})] collection of manifolds \be  \label{3.17}
\sigma_2=\left\{\Sigma^2_k|k\in K_2\right\} \ee containing prime
compact closed three-manifolds and the sphere $S^3$ (as a neutral
element). ($K_2$ is a finite set of indices.) Using this
collection and the partition (\ref{3.6}), we construct $M_2$ in an
analogy with (\ref{3.4}) as a union of connected sums,
$$
M_2=\bigcup_{{\scriptsize\begin{array}{c}
i_0\in I_0\\
i_1\in I_1(i_0)
\end{array}}}
\left(A_{i_0i_1}\#\Sigma^2_{k(i_0,i_1)}
\right), ~ ~ k(i_0,i_1)\in K_2.
$$

\underline{Observation 3.3.} Note again that in fact we define a
family of 2c-manifolds depending of two integer-valued functions,
$k(i_0)$ ({\em cf.} Observation 3.1) and $k(i_0,i_1)$. Thus
instead of $M_2$ one should write $M_{2,k(i_0),k(i_0,i_1)}$. When
the both functions are fixed, we shall use the abbreviation $M_2$.

Applying this procedure repeatedly, we come to the hc-manifold
$M_h$ (more exactly, to a family of hc-manifolds, see the
Observation 3.3). We give here the respective expressions.

Introduce a collection \be  \label{3.18}
\sigma_h=\left\{\Sigma^h_k|k\in K_h\right\} \ee which includes
prime compact closed manifolds, as well as $S^3$. Then $M_h$ is
constructed with help of the connected sum operation applied to
each set of the partition $\alpha_{h-1}$, \be  \label{3.19}
M_h=\bigcup_{{\scriptsize\begin{array}{c}
i_p\in I_p(i_0,\dots,i_{p-1})\\
p=0,\dots,h-1
\end{array}}}
\left(A_{i_0,\dots,i_{h-1}}\#\Sigma^h_{k(i_0,\dots,i_{h-1})}
\right) \ee where $k(i_0,\dots,i_{h-1})$ is a fixed integer-valued
function with values in $K_h$. We automatically come to the
canonical (though not nice) partition \be  \label{3.20}
\begin{array}{c}
\alpha'_{h-1}=\left\{A'_{i_0,\dots,i_{h-1}}=
A_{i_0,\dots,i_{h-1}}\#\Sigma^h_{k(i_0,\dots,i_{h-1})}|
\right.\\
\left.i_p\in I^{\phantom{|}}_p(i_0,\dots,i_{p-1}), ~ p=0,
\dots,h-1,k(i_0,\dots,i_{h-1})\in K_h\right\}.
\end{array}
\ee Here $I_p(i_0,\dots,i_{p-1})$ is a finite collection of
indices which enumerates the sets $A_{i_0,\dots,i_p}\in\alpha_p$
being subsets of $A'_{i_0,\dots,i_{p-1}}$, {\em i.e.} \be
\label{3.21} A'_{i_0,\dots,i_{p-1}}=\bigcup_{i_p\in
I_p(i_0,\dots,i_{p-1})} A_{i_0,\dots,i_p}. \ee We have again the
isomorphism $M_{{h-1},\alpha_{h-1}}\cong M_{h,\alpha'_{h-1}}$ of
$T_0$-discrete spaces corresponding to different c-manifolds
$M_{h-1}$ and $M_h$.

To complete the construction process, one has to introduce a nice
canonical partition \be  \label{3.22}
\alpha_{h}=\left\{A_{i_0,\dots,i_h}|i_p\in I_p(i_0,\dots,i_{p-1}),
~ p=0,\dots,h\right\} \ee of the manifold $M_h$; $\alpha_h$ is a
Lebesgue refinement of the partition $\alpha'_{h-1}$, which
satisfy the condition \be  \label{3.23}
d(\alpha_h)\leqslant\min\left(\frac{1}{2}d(\alpha'_{h-1}),
\frac{1}{2}l(\alpha'_{h-1})\right) \ee ({\em cf.} the Observation
3.2).

The homomorphism \be  \label{3.24}
\omega^{\alpha_h}_{\alpha_{h-1}}:=\omega^{\alpha_h}_{
\alpha'_{h-1}}:M_{h,\alpha_h}\rightarrow M_{h,\alpha'_{h-1}} \cong
M_{{h-1},\alpha_{h-1}} \ee again gives the discrete description of
the topology change between the manifolds $M_h$ and $M_{h-1}$.
$T_0$-discrete space \be  \label{3.25} M^{{\rm
cr}}_{h-1,h}:=M_{h,\alpha'_{h-1}} \cong M_{{h-1},\alpha_{h-1}} \ee
represents the critical level at which occurs the bifurcation of
bispectrum \be  \label{3.26}
\begin{array}{r}
\left\{M_{h-1,\alpha_{h-1,i}},\omega^{\alpha_{h-1,i+1}}_{
\alpha_{h-1,i}}| i\in\overline{p_{h-1},\infty}\right\},\\
\left\{M_{h,\alpha_{h,i}},\omega^{\alpha_{h,i+1}}_{
\alpha_{h,i}}| i\in\overline{p_{h},\infty}\right\}^{
\phantom{|}}
\end{array}
\ee in the same sense as for the bispectra (\ref{3.16}).

\subsection{Multispectra, superspectra, and their
many-world interpretation}  \label{s3.3}

In the preceding Subsection there has been constructed a sequence
of topology changes between compact manifolds $M_{n+1}\rightarrow
M_n$ ($n=0,1,\dots,h-1$) in terms of a sequence of homomorphisms
between $T_0$ Alexandroff spaces \be  \label{3.27}
\omega^{\alpha_{n+1}}_{\alpha_n}:M_{n+1,\alpha_{n+1}} \rightarrow
M_{n,\alpha_n} \ee [see, {\em e.g.}, (\ref{3.24})] where
$\alpha_n$ and $\alpha_{n+1}$ are nice canonical partitions of
manifolds $M_n$ and $M_{n+1}$ respectively, which satisfy the
Lebesgue refinement condition \be  \label{3.28}
d(\alpha_{n+1})\leqslant\min\left(\frac{1}{2}d(\alpha'_n),
\frac{1}{2}l(\alpha'_n)\right) \ee (see Observation 3.2). We have
supposed herewith that all integer-valued functions
$k(i_0,\dots,i_n)$ are fixed. Then it was possible to describe
these topology changes in terms of bifurcations of ($n,n+1$)-level
bispectra, \be  \label{3.29}
\begin{array}{r}
\left\{M_{n,\alpha_{n,i}},\omega^{\alpha_{n,i+1}}_{
\alpha_{n,i}}| i\in\overline{p_{n},\infty}\right\},\\
\left\{M_{n+1,\alpha_{n+1,i}},\omega^{\alpha_{n+1,i+1}}_{
\alpha_{n+1,i}}| i\in\overline{p_{n+1},\infty}\right\},
\end{array}
\ee
where $p_{n+1}>p_n$ for all $n=0,1,\dots,h-1$.

In this Subsection we lift the restriction concerning the
fixedness of functions $k(i_0,\dots,i_n)$ in the definition of
manifolds of all levels, thus coming to the concepts of
multispectrum and superspectrum of compact manifolds.

Passing to the construction of a multispectrum it is worth
remembering that (see the Observation 3.1) the set $J_1$ of the
mappings $k:I_0\rightarrow K_1$ has the cardinality
$$
|J_1|=|K_1|^{|I_0|}
$$
where $|I_0|$ is the number of sets in the nice canonical
partition $\alpha_0=\{A_{i_0}|i_0\in I_0\}$ of the connected
0c-manifold $M_0$, while $|K_1|$ is the number of elements in the
collection $\sigma_1$ (\ref{3.3}). Thus when all possible
integer-valued functions $k(i_0)\in J_1$ are admissible, the
compact manifold of the first level (1c-manifold) becomes {\em
multicomponent} one, \be  \label{3.30} M^{{\rm
mc}}_1=\bigsqcup_{k(i_0)\in J_1}\bigcup_{i_0\in
I_0}\left(A_{i_0}\#\Sigma^1_{k(i_0)}\right) \ee ($\bigsqcup$ is a
disjoint union) with the number $|J_1|$ of connected components
such as \be  \label{3.31} M_{1,k(i_0)}=\bigcup_{i_0\in
I_0}\left(A_{i_0}\# \Sigma^1_{k(i_0)}\right)\equiv\bigcup_{i_0\in
I_0} A'_{i_0,k(i_0)} \ee [{\em cf.} (\ref{3.4})]. We automatically
come to the canonical (though not nice) partition \be
\label{3.32} \alpha'_{0,k(i_0)}=\left\{A'_{i_0,k(i_0)} |i_0\in
I_0\right\} \ee of any component $M_{1,k(i_0)}$ of the manifold
$M^{{\rm mc}}_1$. Then we have the homeomorphism \be  \label{3.33}
M_{0,\alpha_0}\cong M_{1,k(i_0),\alpha'_0} \ee of $T_0$-discreet
spaces, like in the Subsection \ref{s3.2}. Here $M_{0,\alpha_0}$
is a $T_0$-discrete space corresponding to the partition
$\alpha_0$ of the manifold $M_0$, while $M_{1,k(i_0),\alpha'_0}$
is a $T_0$-discrete space corresponding to the partition
$$
\alpha'_0=\left\{A'_{i_0,k(i_0)}|i_0\in I_0,
k(i_0)\in J_1\right\}
$$
of the manifold $M^{{\rm mc}}_1$ restricted onto the component
$M_{1,k(i_0)}$. Due to this homeomorphism there is the covering
mapping \be  \label{3.34} \pi_0:\bigsqcup_{k(i_0)\in
J_1}M_{1,k(i_0),\alpha'_0} \rightarrow M_{0,\alpha_0} \ee such
that its restriction onto one component $M_{1,k(i_0),\alpha'_0}$
is an identical homeomorphism. This covering mapping in fact
involves a possibility to describe the topology change between the
manifolds $M_0$ and $M^{{\rm mc}}_1$. In order to give this
description strictly in terms of inverse spectra, we introduce the
following definitions analogous to those given in the Subsection
\ref{s3.2}.

Consider two sequences of canonical partitions $\{\alpha_{0,i}\}$
and $\{\alpha_{1,i}\}$ of the manifolds $M_0$ and $M^{{\rm mc}}_1$
respectively. Here $i\in\overline{p_0,\infty}$. For $i=p_0$ we put
$\alpha_{0,p_0}=\{M_0\}$ and $\alpha_{1,p_0}=\{M^{{\rm mc}}_1\}$,
thus the respective one-point spaces are homeomorphic, \be
\label{3.35} M_{0,\alpha_{0,p_0}}\cong M_{1,\alpha_{1,p_0}}. \ee
For some $i=p_1>p_0$, set $\alpha_{0,p_1}=\alpha_0$ and
$\alpha_{1,p_1}=\alpha'_0$. Thus due to (\ref{3.33}) the
homeomorphism \be  \label{3.36} M_{0,\alpha_{0,p_1}}\cong
M_{1,k(i_0),\alpha_{1,p_1}} \ee holds for any $k(i_0)\in J_1$.
Then similar to (\ref{3.15}) one can introduce the pair of inverse
spectra \be  \label{3.37}
\left\{M_{0,\alpha_{0,i}},\omega^{\alpha_{0,i+1}}_{
\alpha_{0,i}}|i\in\overline{p_0,\infty}\right\}, ~ ~
\left\{M^{{\rm mc}}_{1,\alpha_{1,i}},\omega^{\alpha_{1,
i+1}}_{\alpha_{1,i}}|i\in\overline{p_0,\infty}\right\}. \ee

It is worth emphasizing that for $i\in\overline{p_0,p_1}$, all
components $M_{1,k(i_0),\alpha_{1,i}}$ of $M^{{\rm
mc}}_{1,\alpha_{1,i}}$ may be defined in such a manner that they
will be homeomorphic to $M_{0,\alpha_{0,i}}$, so that they can be
identified among themselves as well as with $M_{0,\alpha_{0,i}}$
[see (\ref{3.36}) and the reasonings concerning the enlargement of
partitions in the Subsection \ref{s3.2} with respect to
(\ref{3.13}) and (\ref{3.14})]. Besides, $M_{1,k(i_0)}=M_0$ for
$k(i_0)\equiv 0$ since in this case the expression (\ref{3.31})
takes the form
$$
M_{1,k(i_0)}|_{k(i_0)\equiv 0}=\bigcup_{i_0\in I_0}
\left(A_{i_0}\# S^3\right)=\bigcup_{i_0\in I_0}
A_{i_0}=M_0.
$$
Thus $M_0$ enters $M^{{\rm mc}}_1$ as one of its components.
Consequently, the first spectrum in the pair (\ref{3.37}) can be
restricted to the interval $i\in\overline{p_0,p_1}$, while the
second one will begin at $i=p_1$, giving the following definition
to the (0,1)-{\em level multispectrum}: \be  \label{3.38}
\left\{M_{0,\alpha_{0,i}},\omega^{\alpha_{0,i+1}}_{
\alpha_{0,i}}|i\in\overline{p_0,p_1}\right\}, ~ ~ \left\{M^{{\rm
mc}}_{1,\alpha_{1,i}},\omega^{\alpha_{1,
i+1}}_{\alpha_{1,i}}|i\in\overline{p_1,\infty}\right\}. \ee This
term is justified by the fact that (\ref{3.38}) describes the
multifurcation \be  \label{3.39} M_{0,\alpha_0}\rightarrow
\bigsqcup_{k(i_0)\in J_1} M_{1,k(i_0),\alpha_{1,p_1}} \ee at the
critical level (\ref{3.36}), {\em i.e.} for $i= p_1$, $M^{{\rm
cr}}_{01}:=M_{0,\alpha_{0,p_1}}$.

Further we introduce the nice canonical partition
$$
\alpha_1=\{A_{i_0i_1,k(i_0)}|i_0\in I_0,i_1\in I_1
(i_0,k(i_0)),k(i_0)\in J_1\}
$$
of the multicomponent 1c-manifold $M^{{\rm mc}}_1$. Note that in
this case one ought to explicitly write out the collection of
indices $I_1(i_0,k(i_0))$ as a function not only of $i_0$ but also
of the integer-valued function $k(i_0)\in J_1$ which enumerates
the components of $M^{{\rm mc}}_1$. Thus instead of (\ref{3.7}) we
have \be  \label{3.40} A'_{i_0,k(i_0)}=\bigcup_{i_1\in
I_1(i_0,k(i_0))}A_{i_0 i_1,k(i_0)}. \ee

Note that by the transition to the manifolds of the subsequent
levels, the number of connected components increases dramatically.
Since bulkiness of the expressions increases similarly, we shall
confine ourselves to the second-level manifold $M^{{\rm mc}}_2$
being again defined via the collection of the compacts $\sigma_2$
(\ref{3.17}): \be  \label{3.41}
\begin{array}{rl} M^{{\rm mc}}_2 &
=\displaystyle\bigsqcup_{{\scriptsize\begin{array}{c}
k(i_0,i_1)\in J_2(k(i_0))\\
k(i_0)\in J_1
\end{array}}}\displaystyle
\bigcup_{{\scriptsize\begin{array}{c}
i_1\in I_1(i_0)\\
i_0\in I_0
\end{array}}}
\left(A_{i_0i_1,k(i_0)}\#\Sigma^2_{k(i_0,i_1)}
\right)\\
~ & =\displaystyle\bigsqcup_{{\scriptsize
\begin{array}{c}
k(i_0,i_1)\in J_2(k(i_0))\\
k(i_0)\in J_1
\end{array}}}
M_{2,k(i_0),k(i_0,i_1)}.
\end{array}
\ee Here $J_2(k(i_0))$ is the set of components of the manifold
$M^{{\rm mc}}_2$ arising from one component $M_{1,k(i_0)}$ of the
manifold $M^{{\rm mc}}_1$ due to the arbitrariness of the
functions $k(i_0,i_1)$ [for any fixed function $k(i_0)$]. In other
words, $J_2(k(i_0))$ is a set of mappings \be  \label{3.42}
k:I_1(k(i_0))\rightarrow K_2 \ee where $I_1(k(i_0))$ is the set of
elements of the partition $\alpha_1$ pertaining to component
$M_{1,k(i_0)}$ of the manifold $M^{{\rm mc}}_1$, {\em i.e.} \be
\label{3.43} I_1(k(i_0))=\bigcup_{i_0\in I_0}I_1(i_0,k(i_0)) \ee
[{\em cf.} the expression (\ref{3.40})].

Then the number of components of the manifold $M^{{\rm mc}}_2$
arising from one component $M_{1,k(i_0)}$, is \be  \label{3.44}
|J_2(k(i_0))|=|K_2|^{|I_1(k(i_0))|}, \ee while the total amount of
components of $M^{{\rm mc}}_2$ is \be  \label{3.45}
|J_2|=\sum_{k(i_0)\in J_1}|J_2(k(i_0))|. \ee (Remember that $|A|$
means the cardinality of the set $A$.)

The expression (\ref{3.41}) contains the definition of a canonical
(but not nice) partition \be  \label{3.46}
\begin{array}{l} \displaystyle\alpha'_1=\left.
\left\{A'_{i_0i_1,k(i_0),k(i_0,i_1)}=
A_{i_0i_1,k(i_0)}\#\Sigma^2_{k(i_0,i_1)}\right|\right.\\
\displaystyle\left.i_0\in I_0,i_1\in I_1(i_0,k(i_0)),k(i_0)\in
J_1,k(i_0,i_1)\in J^{\phantom{|^o}}_2(k(i_0))\right\}  \end{array}
\ee of $M^{{\rm mc}}_2$; by analogy with (\ref{3.33}), exist
homeomorphisms
$$
M_{1,k(i_0),\alpha_1}\cong M_{2,k(i_0),k(i_0,i_1),
\alpha'_1}, ~ k(i_0)\forall\in J_1,k(i_0,i_1)\forall\in J_2(k(i_0))
$$
between connected components of $T_0$-discrete spaces $M^{{\rm
mc}}_{1,\alpha_1}$ and $M^{{\rm mc}}_{2, \alpha'_1}$ respectively.
Then, using the prescription given in (\ref{3.38}) for the
(0,1)-level multispectrum, one can determine the (1,2)-level
multispectrum as \be  \label{3.47} \left\{M^{{\rm
mc}}_{1,\alpha_{1,i}},\omega^{\alpha_{1,
i+1}}_{\alpha_{1,i}}|i\in\overline{p_1,p_2}\right\}, ~ ~
\left\{M^{{\rm mc}}_{2,\alpha_{2,i}},\omega^{\alpha_{2,
i+1}}_{\alpha_{2,i}}|i\in\overline{p_2,\infty}\right\}, \ee as
well as $(n,n+1)$-level multispectra for any $n=0,1, \dots,h-1$ as
\be  \label{3.48} \left\{M^{{\rm
mc}}_{n,\alpha_{n,i}},\omega^{\alpha_{n,
i+1}}_{\alpha_{n,i}}|i\in\overline{p_n,p_{n+1}}\right\}, ~ ~
\left\{M^{{\rm mc}}_{n+1,\alpha_{n+1,i}},\omega^{\alpha_{n+1,
i+1}}_{\alpha_{n+1,i}}|i\in\overline{p_{n+1},\infty}\right\}. \ee
Here $p_{n+1}>p_n$ and in contrast to the bispectrum (\ref{3.29}),
the first inverse spectrum in (\ref{3.48}) can be taken as finite,
since for $i>p_{n+1}$ all $T_0$-discrete spaces of the first
spectrum are contained in the $T_0$-discrete spaces of the second
one, \be  \label{3.49} M^{{\rm mc}}_{n,\alpha_{n,i}}\subset
M^{{\rm mc}}_{n+1, \alpha_{n+1,i}}. \ee The upper limit of the
multispectrum (\ref{3.48}) is the multicomponent compact manifold
$M^{{\rm mc}}_{n+1}$ which contains subspaces forming a chain of
inclusions,
$$
M^{{\rm mc}}_{n+1}\supset M^{{\rm mc}}_n\supset\dots
\supset M^{{\rm mc}}_1\supset M_0.
$$

The multispectrum (\ref{3.48}) describes multifurcations \be
\label{3.50} \begin{array}{r}
M_{n,k(i_0),\dots,k(i_0,\dots,i_{n-1}),\alpha_{n,p_{n+1}}}
\rightarrow\\  ~~~~\\
\displaystyle\bigsqcup_{{\scriptsize\begin{array}{c}
k(i_0),\dots,i_n)\in\\  J_{n+1}(k(i_0),
\dots k(i_0,\dots,i_{n-1})) \end{array}}}\mbox{\hspace*{-1.cm}}
M_{n+1,k(i_0),\dots,k(i_0,\dots,i_n),\alpha_{n+1,p_{n+1}}}
\end{array}
\ee at the $(n+1)$th critical level determined by the
multicomponent $T_0$-discrete space \be  \label{3.51} M^{{\rm
cr}}_{n,n+1}:=M^{{\rm mc}}_{n,\alpha_{n,p_{n+1}}}. \ee

The union of all multispectra (\ref{3.48}) for $n=0,\dots,h-1$ we
shall call the $h$-level {\em superspectrum} in an analogy with
the Wheeler--DeWitt superspace. This analogy will follow from the
interpretation of multi- and superspectra to which we turn now.

On a multicomponent manifold $M^{{\rm mc}}_n$ being one of
subspaces of the upper inverse limit of the multispectrum
(\ref{3.48}), let us introduce a partition $\alpha_{n,i}$ where
$i\in\overline{p_n,p_{n+1}}$. The diameter of this partition we
denote as $d_{n,i}$. One may now say that to this partition
corresponds a {\em system of observers} (forming a discrete
reference frame introduced in the Subsection \ref{s2.4}) which
realize measurements at the scales $d_{n,i}$ in the interval \be
\label{3.52} d(\alpha_{n+1,p_{n+1}})<d_{n,i}<d(\alpha_{n,p_n}).
\ee

In this case we shall say that a subsystem of mutually
interconnected (via chains of the sets of $\alpha_{n,i}$)
observers may identify itself in one of the connected components
of the $T_0$-discrete space $M^{{\rm mc}}_{n,\alpha_{n,i}}$ of the
$(n,n+1)$-level multispectrum (\ref{3.48}), {\em e.g.}, in the
component \be  \label{3.53}
M_{n,k(i_0),\dots,k(i_0,\dots,i_{n-1}),\alpha_{n,i}}. \ee It is
natural to question of the probability to occur in one or another
component of the space $M^{{\rm mc} }_{n,\alpha_{n,i}}$. From the
viewpoint of the many-world interpretation of quantum cosmology,
the probability amplitude to occur in the universe (\ref{3.53})
should be determined by a certain wave function \be  \label{3.54}
\Psi(M_{n,k(i_0),\dots,k(i_0,\dots,i_{n-1}),\alpha_{n,i}}). \ee
The domain of definition of this probability amplitude function is
the set of connected components of $T_0$-discrete spaces forming
the $h$-level superspectrum \be  \label{3.55}
\begin{array}{rl}
{\rm DSS}= & \left\{M_{n,k(i_0),\dots,k(i_0,\dots,
i_{n-1}),\alpha_{n,i}}|\right.\\
~ & k(i_0,\dots,i_s)\in J_{s+1}(k(i_0),\dots,
k(i_0,\dots,i_{s-1})),\\
~ & \left.s\in\overline{0,n-1},
i\in\overline{p_n,p_{n+1}},n\in\overline{0,h-1}\right\}.
\end{array}
\ee Namely this set we shall call the {\em discrete superspace}
(DSS) which we will treat as an analogue (in our model) of the
Wheeler--DeWitt superspace (Wheeler, 1964; DeWitt, 1967).
Performing measurements in a discrete reference frame, say, that
which corresponds to the partition $\alpha_{n,i}$ [{\em i.e.}, on
the scales (\ref{3.52})], the observers merely find out in which
connected component (\ref{3.53}) ({\em i.e.} at which point of the
DSS) they are localized. This represents just an example of the
many-world interpretation of the wave function (Everett, 1957;
DeWitt and Graham, 1973). Executing more refined measurements,
{\em e.g.}, on the scales \be  \label{3.56}
d(\alpha_{n+2,p_{n+2}})<d_{n+1,i}<d(\alpha_{n+1,p_{n+1}}), \ee the
observers occur in a wider set of connected components forming
$T_0$-discrete space $M^{{\rm mc}}_{n+1, \alpha_{n+1,i}}$. This is
related to the presence of the critical level (\ref{3.51}) at
which the phase transition of an increase of the number of the
multiconnected-space components, (\ref{3.50}), occurs. Hence in
our model a certain kind of Heisenberg's uncertainty principle
holds: the finer scales at which observers perform the
measurements ({\em i.e.}, the more refined the canonical
partition), the wider the ensemble of the connected components of
$T_0$-discrete spaces in which they can detect themselves. However
due to validity of the inclusion condition (\ref{3.49}), there is
a possibility for observers to find themselves in the former
connected component while performing measurements with the
accuracy (\ref{3.56}) greater than that of (\ref{3.52}).

Of course, the problem of evaluation of the wave function
(\ref{3.54}) depends on construction of some topological analogue
of the Wheeler--DeWitt quantum geometrodynamics; it could be
provisionally called discrete quantum topodynamics. The
topodynamical analogue of the Wheeler--DeWitt equation is still
unknown, but for the role of an analogue of the superspace, we
propose the discrete superspace DSS (\ref{3.55}) as the domain of
definition of topodynamical wave functions.

\section{Concluding remarks}  \label{s4}

In this paper the following three results may be emphasized.

1. On the basis of Alexandroff's mathematical constructions and
under the influence of Sorkin's physical ideas, we have modelled
the discrete spacetime via the proper inverse spectrum $S_{{\rm
pr}}$ relating the cosmological time arrow in the expanding
universe to the refinement of canonical partitions $\{\alpha\}$ of
the continuous compact three-dimensional space $X$ homeomorphic to
the upper inverse limit $\hat{S}_{{\rm pr}}$ of this spectrum.
However this limiting compact space $X\cong\hat{S}_{{\rm pr}}$ is
never realized in the course of the discrete cosmological
evolution although at a sufficiently fine partition $\alpha$, the
corresponding $T_0$-discrete space $X_\alpha$ is a good
approximation of the compact $X$. [When the measurements are
performed with an uncertainty $\Delta d$ much greater than the
diameter $d(\alpha)$ of the partition $\alpha$ ($\Delta d\gg
d(\alpha)$), the discrete space $X_\alpha$ will be experimentally
indistinguishable from the continuous space $X\cong\hat{S}_{{\rm
pr}}$.] Moreover the spectral evolution of $T_0$-discrete spaces
automatically suggests the existence of $T_0$-discrete
one-dimensional time being not specially postulated from the
beginning.

It is obvious that in our model the problem of initial
cosmological singularity is automatically lifted, since the
initial one-point space naturally pertains to the same class of
$T_0$-discrete spaces to which all other spatial discrete sections
pertain (in contrast to the Friedmann--Robertson--Walker
cosmological models), see Subsection \ref{s2.4}, (\ref{2.16}) and
Subsection \ref{s3.2}.

Evolution of the universe is described by the sequence of
topological changes (\ref{2.20}); this corresponds to the
transition from the partition $\alpha_i$ to the next more refined
one, $\alpha_{i+1}$. It is natural to associate the `discrete time
quantum' or `elementary topological change' to the act of
subdivision of only one set (element) of the partition $\alpha_i$
into two sets. When we have, say, $10^{360}$ elements of the
partition $\alpha_i$ [{\em cf}. (Rideout and Sorkin, 2001)], the
evolution described by the sequence of elementary topological
changes $\left(\omega^{\alpha_{i+1}}_{\alpha_i}
\right)^{-1}:X_{\alpha_i}\rightarrow X_{\alpha_{i+1}}$ can be
considered as continuous from the viewpoint of an observer with
measuring devices having the resolving power much more coarse than
$10^{-360}$ (in the sense that in the whole discrete universe the
number of elements has increased from $10^{360}$ to $10^{360}+1$).
Moreover, on the proper inverse spectrum $S_{{\rm pr}}$ it turned
out to be possible to define such a relation of partial ordering
(Subsection \ref{s2.5}) that has the basic properties of the
causal order in the pseudo-Riemannian spacetime [transitivity and
mutual compatibility of the sets pertaining to causal past and
causal future (Propositions 2.1 and 2.2)], however also bearing
certain quasi-quantum features [{\em e.g.} dependence of the
causal past and future on the choice of discrete reference frame
at the near-discreteness, say, Planck, scales (Observation 2.4)].

2. Our interpretation of inverse spectra of the three-dimensional
Alexandroff spaces permitted to represent topological changes
between three-mani\-folds in terms of bispectra (\ref{3.29}). Note
that in accordance with the results of Kneser (1929) and Milnor
(1962) this approach permits to describe creation of any
3-manifold $M$ as a result of a succession of topological
transitions from an initial manifold which is homeomorphic to the
sphere $S^3$.

3. Both our the considerations and results obtained here, are
bearing essentially classical and kinematical character (though
they possess certain quasi-quantum features, see, {\em e.g.},
Subsections \ref{s2.4} and \ref{s2.5}). However, having in mind
quantization and the necessity to introduce dynamics, we have
described a discrete analogue of the Wheeler--DeWitt superspace,
just having defined the discrete superspace DSS (\ref{3.55}) as
the arena of the future topodynamics (an analogue of the
Wheeler--DeWitt geometrodynamics). Already at this qualitative
stage we were able to formulate a topological counterpart of the
many-worlds interpretation of the quantum wave function
(\ref{3.54}), as well as an analogue of Heisenberg's uncertainty
relation.

Alternatively, it is possible that introduction of dynamics in our
model could occur similarly to the sequential growth dynamics
(Rideout and Sorkin, 2000). This approach may be adapted to our
model [in the $T_0$-discrete spacetime sandwich construction
described in Subsection \ref{s2.4} (3)].

The algebraic approach to quantum topodynamics based on
constructing the direct spectrum of the \v{C}ech cohomology groups
(Bott and Tu, 1982; Raptis, 2000b; Mallios and Raptis 2001), also
seems to be promising. Then, using operators of coboundary and
homotopy [possessing interesting anticommutation properties
(Teleman, 1964)] which act on \v{C}ech cochain complexes (over
nerves of coverings with values in presheaves of certain algebraic
structures), one could expect to come to an analogue of the
many-particle Fock representation. Here the part of independent
(mutually non-interacting) topological excitations (excitops) of
the $(n+1)$st level, would be played by the prime closed manifolds
from the collection $\sigma_{n+1}=\left\{\Sigma^{n+1}_k|k\in
K_{n+1} \right\}$, see Subsections \ref{s3.2}, \ref{s3.3}. (The
question still remains open which prime closed manifolds form this
collection.) These excitations are determined on the $n$th level
manifold $M_n$ as a classical background. Here the hierarchy of
scales (\ref{3.28}) may be of importance, the hierarchy being
related to the fundamental role of the Lebesgue numbers in
constructing the being ever refined systems of open coverings and
canonical partitions.

For us it was a source of inspiration and moral support to read
and reread the wonderful Richard Courant Lecture in Mathematical
Sciences by Eugene P. Wigner (1960) so masterfully summarized in
his own words: ``The miracle of the appropriateness of the
language of mathematics for the formulation of the laws of physics
is a wonderful gift which we neither understand nor deserve. We
should be grateful for it and hope that it will remain valid in
future research and that it will extend, for better or for worse,
to our pleasure even though perhaps also to our bafflement, to
wide branches of learning.'' In any case, with all future
corrections which life will give to our conclusions in this
article, we are sure that the most abstract branch of mathematics
touched by physicists until now, the topology, is becoming more
and more accessible for a fruitful application in fundamental
physics, opening new horizons in the organic synthesis of
relativity and quantum.

~~~~~~~~~~
~~~~~~~~~~~~~

\section*{REFERENCES}
~~~~~~~~

Alexandroff, P.S. (1929). {\em Annalen der Mathematik} {\bf 30},
101. In German.

Alexandroff, P.S.  (1937). {\em Matematicheskii Sbornik} {\bf 2},
501. In German.

Alexandroff, P.S.  (1947). {\em Uspekhi Matem. Nauk} {\bf 2(17)},
5. In Russian.

Alexandrov, P.S. (1998). {\em Combinatorial topology}, Dover, New
York.

Antonov, V.I., Efremov, V.N., and Vladimirov, Yu.S. (1978). {\it
General Relativity and Gravitation} {\bf 9}, 9.

Arenas, F.G. (1997). {\it Some results on Alexandroff spaces},
Preprint, Topology Atlas, http://at.yorku.ca/p/a/a/k/12.htm .

Arenas, F.G. (1999). {\em Acta Mathematica Univ. Comenian.}
(N.S.), {\bf 68}, No. 1, 17.

Ashtekar, A. (1991). {\em Lectures on non-perturbative canonical
gravity (Notes prepared in collaboration with R. Tate)}, Advanced
Series in Astrophysics and Cosmology, {\bf 6}, World Scientific,
Singapore.

Baierline, R.F., Sharp, D.H., and Wheeler, J.A. (1962). {\em
Physical Review} {\bf 126}, 1864.

Bohm, D. (1965). {\em The Special Theory of Relativity}, W.A.
Benjamin, New York.

Bohr, N., and Rosenfeld, L. (1933). {\em Det Kgl. Videnskabernes
Selskab. Matematisk-fysiske Meddelelser} {\bf 12}, No. 8. In
German.

Boltyansky, V.G. (1951). {\em Uspekhi Matematicheskikh Nauk} {\bf
6}, No. 3, 99. In Russian.

Bombelli, L., Lee, J., Mayer, D., and Sorkin, R.D. (1987). {\em
Physical Review Letters} {\bf 59}, 521.

Bott, R., and Tu, L.W. (1982). {\em Differential forms in
algebraic topology,} Springer-Verlag, New York.

DeWitt, B.S. (1967). {\it Physical Review} {\bf 160}, 1113.

DeWitt, B.S., and Graham, N. (1973). {\em Many-world
interpretation of quantum mechanics}, Princeton University Press,
Princeton, N.J.

Everett, H. (1957). {\it Reviews of Modern Physics} {\bf 29}, 454.

Finkelstein, D. (1969). {\em Physical Review} {\bf 184}, 1261.

Finkelstein, D. (1988). {\em International Journal of Theoretical
Physics} {\bf 27}, 473.

Fomenko, A.T., and Matveev, S.V. (1991). {\em Algorithmic and
computer methods for three-manifolds,} Kluwer Acad. Publ., London.

Hocking, J.G. and Young, G.S. (1988). {\em Topology,} Dover, New
York.

Isham, Ch. (1989). {\em Classical and Quantum Gravity} {\bf 6},
1509.

Kneser, H. (1929). {\em Jahresbericht der Deutschen Mathematischen
Vereinigung} {\bf 38}, 248.

Kodama, Y. (1958). {\em Journal of the Mathematical Society of
Japan} {\bf 10}, 380.

Mallios, A., and Raptis, I. (2001). {\em International Journal of
Theoretical Physics} {\bf 40}, 1241.

Milnor, J. (1962). {\em American Journal of Mathematics} {\bf 84},
1.

Misner, C.W., Thorne, K.S. and Wheeler, J.A. (1973). {\em
Gravitation}, W.H. Free\-man, San Francisco.

Mitskievich, N.V. (1996) {\em Relativistic Physics in Arbitrary
Reference Frames}, arXiv gr-qc/9606051.

Raptis, I. (2000a). {\em International Journal of Theoretical
Physics} {\bf 39}, 1233.

Raptis, I. (2000b). {\it International Journal of Theoretical
Physics} {\bf 39}, 1703.

Raptis, I., and Zapatrin, R.R. (2001). {\em Classical and Quantum
Gravity} {\bf 20}, 4187.

Rideout, D.P., and Sorkin, R.D. (2000). {\em Physical Review} {\bf
D 61}, 024002.

Rideout, D.P., and Sorkin, R.D. (2001). {\em Physical Review} {\bf
D 63}, 104011.

Sorkin, R.D. (1991). {\em International Journal of Theoretical
Physics} {\bf 30}, 923.

Sorkin, R.D. (1995). {\em A specimen of theory construction from
quantum gravity}, arXiv gr-qc/9511063.

Teleman, C. (1964). {\em Elemente de topologie \c{s}i
variet\={a}\c{t}i diferen\c{t}iabile}, Bucure\c{s}ti.

van Dam, H., and Ng, Y.J. (2001) {\em Why $3+1$ metric rather than
$4+0$ or $2+2$?} hep-th/0108067.

Wheeler, J.A. (1964). Geometrodynamics and the issue of the final
state, in {\em Relativity, groups and topology}, DeWitt, C., and
DeWitt, B.S., eds., Gordon and Breach, New York.

Wigner, E.P. (1960). {\em Communications on Pure and Applied
Mathematics} {\bf 13}, 1.

\end{document}